\documentclass[prd,twocolumn,a4paper,superscriptaddress,floatfix,10pt,nofootinbib,longbibliography]{revtex4-2}

\usepackage[T1]{fontenc}
\usepackage[utf8]{inputenc}
\usepackage[english]{babel}
\usepackage{graphicx}
\usepackage{pict2e}
\usepackage{tikz}
\usepackage{amsthm, amssymb}
\usepackage{epstopdf}
\usepackage{float}
\usepackage{subcaption}
\usepackage{braket}
\usepackage{listings}
\usepackage{amsfonts}
\usepackage[normalem]{ulem}
\useunder{\uline}{\ul}{}
\usepackage{multirow}
\usepackage{tikz}
\usepackage{pgfplots}
\usepackage{textgreek}
\pgfplotsset{/pgf/number format/use comma,compat=newest}
\usepackage{siunitx}
\usepackage{ae}
\usepackage{bm}

\usepackage{amsmath}
\usepackage{amsfonts}
\usepackage{amssymb}
\usepackage{graphicx}
\usepackage[titletoc]{appendix}
\usepackage{color}
\usepackage{hyperref}
\usepackage{cleveref}

\def\be{\begin{equation}}
\def\ee{\end{equation}}
\def\bea{\begin{eqnarray}}
\def\eea{\end{eqnarray}}

\definecolor{vividviolet}{rgb}{0.62, 0.0, 1.0}
\definecolor{amaranth}{rgb}{0.9, 0.17, 0.31}
\definecolor{palatinateblue}{rgb}{0.15, 0.23, 0.89}
\definecolor{brightpink}{rgb}{1.0, 0.0, 0.5}
\definecolor{cornflowerblue}{rgb}{0.39, 0.58, 0.93}
\definecolor{deepcarminepink}{rgb}{0.94, 0.19, 0.22}
\definecolor{radicalred}{rgb}{1.0, 0.21, 0.37}

\hypersetup{ linktoc=all,
    colorlinks, linkcolor={palatinateblue},
    citecolor={brightpink}, urlcolor={amaranth}
}

\begin{document}

\title{Particle production from  non–minimal coupling in a  symmetry breaking potential transporting vacuum energy}

\author{Alessio Belfiglio}
\email{alessio.belfiglio@unicam.it}
\affiliation{University of Camerino, Via Madonna delle Carceri, Camerino, 62032, Italy.}
\affiliation{Istituto Nazionale di Fisica Nucleare (INFN), Sezione di Perugia, Perugia, 06123, Italy.}

\author{Youri Carloni}
\email{youri.carloni@unicam.it}
\affiliation{University of Camerino, Via Madonna delle Carceri, Camerino, 62032, Italy.}

\author{Orlando Luongo}
\email{orlando.luongo@unicam.it}
\affiliation{University of Camerino, Via Madonna delle Carceri, Camerino, 62032, Italy.}
\affiliation{Istituto Nazionale di Fisica Nucleare (INFN), Sezione di Perugia, Perugia, 06123, Italy.}
\affiliation{Department of Mathematics and Physics, SUNY Polytechnic Institute, Utica, NY 13502, USA.}
\affiliation{INAF - Osservatorio Astronomico di Brera, Milano, Italy.}
\affiliation{Al-Farabi Kazakh National University, Almaty, 050040, Kazakhstan.}

\begin{abstract}
We propose an inflationary scenario where the inflaton field is non-minimally coupled to spacetime curvature and inflation is driven by a vacuum energy symmetry breaking potential without specifying \emph{a priori} whether the inflaton field is small or large. As we incorporate vacuum energy into our analysis, we further explore the implications of a non-zero potential offset within inflationary dynamics. We propose that vacuum energy can transform into particles as a result of the transition triggered by spontaneous symmetry breaking. This entails a vacuum energy cancellation that yields an effective cosmological constant during inflation by virtue of a quasi-de Sitter evolution and shows that vacuum energy contribution can manifest as \emph{geometric particles} produced by inflaton fluctuations, with particular emphasis on super-Hubble modes. We conjecture these particles as \emph{quasi-particles} arising from interaction between the inflaton and spacetime geometry, enhanced by non-minimal coupling. Specifically, we propose that dark matter arises from a pure geometric quasi-particle contribution, quantifying the corresponding dark matter candidate ranges of mass. In this scenario, we further find that a zero potential offset leads to a bare cosmological constant at the end of inflation, while a negative offset would require an additional kinetic (or potential) contribution in order to be fully-canceled. In this regard, we conclude that the scenario of large-field inflation is preferred since it necessitates a more appropriate selection of the offset. Our conclusion is reinforced as small-field inflation would lead to a significant screening of the Newtonian gravitational constant as inflation ends. 

\end{abstract}

\maketitle
\tableofcontents

\section{Introduction}

The very early universe was marked by a period of intense inflation, during which the universe underwent a phase of rapid expansion fueled by an unknown field dubbed \emph{inflaton} \cite{Infl0, tsurev, baurev, Infl1}. Understanding its fundamental properties represents one of the most significant challenges for modern cosmology. Interestingly, inflation is not the only phase in which the universe has undergone accelerated expansion. Indeed, an unknown fluid known as dark energy\footnote{Dark energy is commonly modeled by virtue of barotropic fluids \cite{eos1}, albeit it appears also licit to employ scalar fields or alternatives to Einstein's gravity \cite{eos2}. } is responsible for driving the current universe's accelerated expansion \cite{peebrev, copde}. Dark energy is expected to exhibit exotic properties, generating a negative pressure to counterbalance the action of gravity.

In this respect, the current understanding of the universe's background cosmology is based on the $\Lambda$CDM paradigm \cite{LCDM}, where the cosmological constant, $\Lambda$, dominates over matter and is typically interpreted through quantum fluctuations of vacuum energy \cite{Martin}. However, this model seems to be theoretically incomplete, suffering from a strong fine-tuning problem and several other caveats, among which the coincidence problem and cosmological tensions,  see e.g. \cite{burcc,dp9}. These issues can be thought to collectively come from the so-called \emph{cosmological constant problem}, i.e., the undeniable limitation in reconciling the observed cosmological constant with large theoretical zero-point fluctuations\footnote{The cosmological constant problem has significant implications for theoretical physics, since it involves some aspects of quantum gravity but, at the same time, its effects are present on large scales, thus observational signatures are in principle detectable.}. Solving it would represent a crucial step towards understanding physics beyond the current standard models of cosmology and particle physics.

Within this picture, inflation emerges as an interesting playground where quantum mechanics and gravity are intertwined, with the possibility to detect observable effects arising from primordial fluctuations at current time. For this reason, the inflationary stage represents our starting point in order to address the cosmological constant problem. In principle, the cosmological constant receives contributions from different origins, which cannot be neither cancelled by hand nor ignored within any inflationary or dark energy models \cite{Sakharov, bou, Sola, CCP0, CCP1, CCP2, CCP3}. In this puzzle, \emph{unified dark energy models} acquired great importance as they represent unified scenarios in which dark energy emerges as a consequence of dark matter's existence\footnote{Those models describe both dark energy and dark matter using a single fluid \cite{DF0, DF1}.}, see e.g. \cite{ude1, ude2, ude3, ude4, ude5}. Even though the idea of unifying the dark sector is widely-consolidate, current accelerated phase and inflation are still thought to be distinct scenarios.

In analogy to unified dark energy models, however, there are attempts to unify inflation with current universe speed up into a single theoretical framework in which the fluid filling the universe energy budget modifies its properties as the universe expands, giving rise to both inflaton and dark energy  \cite{DEandInf0, DEandInf1, DEandInf2, DEandInf3}.

In addition, more recently there has been a growing interest in those unified dark energy models addressing the cosmological constant problem by combining inflation with dark energy into a sort of \emph{unified inflationary dark sector paradigm} \cite{gao, lim}. Specifically, as a robust example one can focus on \emph{quasi-quintessence dark fluid},  proposed as a potential solution to eliminate quantum fluctuations of vacuum energy \cite{luoqq}. In this respect, it has been shown that cancelling $\Lambda$ involves the conversion of $\Lambda$-energy excess into particles, possibly identified as dark matter quasi-particles \cite{geocorr, CCPBelfiglio0, CCPBelfiglio1} by means of a geometric mechanism of particle production\footnote{This class of models reconciles inflation with dark energy at later times, providing a bare cosmological constant as a consequence of coupling the inflaton with curvature. Within this picture, but even in a more general sense, the Planck satellite measurements did not exclude \emph{a priori} either small or large fields during inflation \cite{Planck}. Examples of small and large-field models that overcome recent bounds presented in Planck's results are the hilltop and Starobinsky potentials, respectively.} \cite{fri, ces}.

According to these results, it seems that the most prominent inflationary paradigms involve \emph{scalar fields transporting vacuum energy}  \cite{stein, ssb2} that end up into a reheating period, where thermal energy is converted into particles \cite{reh1,reh2,reh3}, with a sufficiently high reheating temperature to generate the observed baryon asymmetry \cite{basym,bario}. Quantum fluctuations have been also recently investigated in the context of scalar field tunneling between two degenerate vacua, showing that the resulting true vacuum state can asymptotically lead to a de Sitter phase, potentially representing dark energy at late times \cite{pla}.

Inspired by unified inflationary scenarios, we explore the implications of a non–minimal Yukawa-like coupling that can potentially address the cosmological constant problem. Specifically, we investigate both the hypotheses of small and large inflaton fields and we consider the universe to undergo a $\Lambda$-dominated phase driving inflation, induced by a symmetry breaking potential through a quasi-de Sitter phase.

We delve into a non-zero potential offset, providing the cosmological constant contribution, during and after inflation. We thus propose a $\Lambda$-cancellation mechanism \emph{occurring during inflation}, where vacuum energy releases its energy to transform into particles,  created from vacuum fluctuations of the inflaton by exploiting field-curvature coupling.

Since these particles derive from an interaction Lagrangian, we consequently conjecture to be particles dressed by the interaction itself, i.e., their nature is revised as \emph{quasi-particles derived from geometry}. Nevertheless, toward this direction, geometric particle production represents an alternative to gravitational particle production from vacuum \cite{gpp1, gpp2}, widely-used to explain dark matter creation during inflation \cite{dmg1, dmg2, dmg3, dmg4, dmg5, dmg6}. We discuss the differences between the two approaches, highlighting that geometric production arises from a perturbative approach. Thus, we first study the dynamics of inflaton fluctuations and then the corresponding spacetime perturbations for small and large fields, investigating how those quasi-particles influence the inflationary dynamics and reduce vacuum energy. To do so, we obtain the scattering $\hat S$ matrix due to the interaction between the field and geometry, within the Dyson approximation, and derive the corresponding probability amplitude for particle production. Afterwards, we demonstrate that geometric production \emph{is not influenced} by the choice of the initial offset, showing that the potential cannot solely cancel out the cosmological constant, that remains therefore not fully-erased. We then derive the particle number density produced at the end of the slow-roll phase,  predicting the corresponding mass limits on geometric quasi-particles that turn out to be heavy fields.  In this respect, we also remark that the presence of non-minimal coupling affects Newton's gravitational constant $G$, modifying its value throughout inflation, which becomes dependent on the field value throughout the inflationary phase. In particular, we will notice that small-field inflation would lead to a significant screening of the Newtonian gravitational constant, after inflation.

Physical consequences on the bare cosmological constant, resulting in the end of inflation, are then discussed. We find that a zero potential offset yields a bare cosmological constant at the end of inflation, while a negative offset would require \emph{a further but constant kinetic or potential contribution} coming from inflation, in order to be erased. In this direction, large-field inflation turns out to be more likely, since it necessitates a more appropriate selection
of the offset. At the same time, the presence of inhomogeneities inevitably breaks the assumption of translational invariance invoked in Weinberg's no-go theorem \cite{CCP0}, allowing for a cancellation mechanism without any fine-tuning of the cosmological constant \cite{nogo, nogoqg}. Finally, comparisons with previous mechanisms of inflationary cancellation are also discussed.

The paper is structured as follows. In Sect. \ref{sezione2}, we investigate the symmetry breaking inflationary phase with the introduction of a symmetry breaking potential, non-minimally coupled with scalar curvature. Implications on the Newton's constant and on fluctuations in the slow-roll regime are discussed. In Sect. \ref{sezione3}, the small-field approach is investigated, focusing on super-Hubble scales. Analogously, in Sect. \ref{sezione4}, the same is developed in the context of large fields. The contribution to dark matter magnitude is explored in Sect. \ref{sezione5} for both small and large fields, while a direct comparison between the two scenarios is reported in Sect. \ref{sezione6}. Finally, in Sect. \ref{sezione7} we report conclusions and perspectives of our work.

\section{Symmetry breaking inflationary phase}\label{sezione2}

We consider a scalar inflaton field non-minimally coupled to the spacetime scalar curvature $R$, with corresponding Lagrangian density
\begin{equation}\label{LagrduringInfl}
    \mathcal{L}=\frac{1}{2}g^{\mu\nu}\partial_{\mu}\phi\partial_{\nu}\phi-V^{\rm eff}\left(\phi,R\right)\,.
\end{equation}
 We underline that only partial derivatives appear in the standard kinetic term for scalar fields, and we set
\begin{equation}\label{effpot}
    V^{\rm eff}\left(\phi,R\right)=V_{0}+\frac{\chi}{4}\left(\phi^{2}-v^{2}\right)^{2}+\frac{1}{2}\xi\phi^{2}R.
\end{equation}
The here prompted potential induces spontaneous symmetry breaking, a mechanism already proposed in single-field inflationary scenarios \cite{ssb1,ssb2,ssb3}. At the same time, the presence of non-minimal coupling is fundamental in order to satisfy the observational constraints provided by Planck measurements \cite{Planck}. In Eq. \eqref{effpot} the quantity $V_0$ denotes the classical offset of the potential, while $v$ is the vacuum expectation value of the inflaton field.  The Lagrangian \eqref{LagrduringInfl} provides inflationary slow-roll solutions both in the limit of small and large fields, i.e., the scalar field may evolve to its final value after transition ($\phi=v$) either from $\phi=0$ or from $\phi \gg v$. In both cases, the inflaton potential contributes with an additional term to the overall cosmological constant
\begin{equation}
    \Lambda_{\rm eff}=\Lambda_{B}+V^{\rm eff}\left(\phi_{min}\right), \label{Latot}
\end{equation}
where $\Lambda_{B}$ represents a ``bare" contribution to the cosmological constant and $V^{\rm eff}\left(\phi_{min}\right)$ is the inflaton potential computed in its minimum. At this level, $\Lambda_B$ simply represents an integration constant,  compatible with general covariance. The energy-momentum conservation is unaltered by it \cite{Martin} and so we have no reason to discard it \emph{a priori}. Usually, this term represents the \emph{bare cosmological constant} and, then, ca proper cancellation of $V^{\rm eff}\left(\phi_{min}\right)$ can easily alleviate the fine-tuning associated with the cosmological constant problem, since $\Lambda_B$ is not related to any quantum contribution. At the same time, the choice of the offset $V_0$ is also fundamental in determining the total amount of vacuum energy before and after the phase transition.

\subsection{Non-minimal coupling and Newton's gravitational constant} \label{Sez2A}

The presence of non-minimal field-curvature coupling in Eq. \eqref{LagrduringInfl} also implies that Newton's gravitational constant $G$  is modified during inflation, and its final value is determined by the field value at the minimum, namely $v$.   During the inflationary phase, we can write the total action as
\begin{equation} \label{totac}
    \mathcal{S}_{tot}=\int d^{4}x\sqrt{-g}\left(-\frac{\Lambda_{\rm eff}}{16\pi G}g_{\mu\nu}-\frac{\xi}{2}R \phi^{2}+\frac{M_{P}^{2}}{16\pi}R+\mathcal{L}_{M}\right),
\end{equation}
where in $\mathcal{L}_{M}$ we included all matter field contributions, which are not relevant for our argument\footnote{During slow-roll, this term simply reduces to the minimally-coupled inflaton Lagrangian density.}. Eq. \eqref{totac} shows that during inflation the effective value of Newton's constant varies, and it crucially depends on the initial value of the inflaton field. At the end of the phase transition, we can assume $\phi \equiv v$, and minimizing the action above we find
\begin{equation} \label{Einsteinmodify2}
G_{\mu\nu}+\frac{\Lambda_{\rm eff}}{1-8\pi G\xi v^{2}}g_{\mu\nu}=-8\pi\tilde{G}T_{\mu\nu},
\end{equation}
with $T_{\mu\nu}$ the energy-momentum tensor. The effective gravitational constant is then
\begin{equation}
  \tilde{G}\equiv \frac{G}{1-8\pi G\xi v^{2}}.
\end{equation}
Thus, we need $1\gg 8\pi G\xi v^{2}$ to hold, in order to preserve the value of Newton's constant and recover the original Einstein field equations, with negligible modifications. We will discuss how non-minimal coupling affects $G$ for both small and large fields later on.

In what follows, we need to better understand the details of the phase transition, focusing in particular on the dynamics of the inflaton fluctuations during slow-roll.


\subsection{Inflaton fluctuations during slow-roll} \label{Sez2B}

The evolution of the inflaton field during slow-roll is usually derived by assuming a flat Friedmann-Robertson-Walker (FRW) background, described by the line element
\be \label{linel}
ds^2=dt^2-a(t)^2 \delta_{ij} dx^i dx^j,
\ee
where $a(t)$ is the scale factor and $t$ denotes cosmic time.
From Eq. \eqref{LagrduringInfl}, we obtain the following equation of motion for the field
\begin{equation}
\ddot{\phi}+3H\dot{\phi}-\frac{\nabla^{2}\phi}{a^{2}}+6\xi \left(\dot{H}+2H^{2}\right)\phi + V\left(\phi\right)_{,\phi}=0,\label{ddotphi}
\end{equation}
where $V$ denotes the above discussed symmetry breaking potential and $R=6\big(\dot{H}+2H^{2}\big)$ is the scalar curvature.

Eq. \eqref{ddotphi} describes the overall dynamics of the inflaton field: it also includes the contribution of its quantum fluctuations, which are expected to be responsible for the formation of large-scale structures in the universe \cite{Infl0, tsurev, baurev} and similarly are the seed of the geometric mechanism of particle production that we are going to discuss.

To investigate quantum fluctuations, for convenience we split the inflaton field by considering fluctuations $\delta \phi({\bf x}, t)$ around its background value:
\begin{equation}
\phi\left(\textbf{x},\tau\right)=\phi_{0}\left(\tau\right)+\delta \phi\left(\textbf{x},\tau\right).\label{totalfieldequation}
\end{equation}
Inflaton fluctuations will, in turn, induce metric perturbations via Einstein's equations. For scalar perturbations, the general FRW perturbed metric reads
\begin{subequations}
    \begin{align}
g_{00}&=a^{2}\left(\tau\right)\left(1+2\Psi\left(\textbf{x},\tau\right)\right), \label{gzero}\\     g_{0i}&=-2a^{2}\left(\tau\right)\partial_{i}B\left(\textbf{x},\tau\right),\label{goffd}\\
g_{ij}&=-a^{2}\left(\tau\right)\left[\left(1-2\Phi\left(\textbf{x},\tau\right)\right)\delta_{ij}+D_{ij}E\left(\textbf{x},\tau\right)\right], \label{gdiag}
    \end{align}
\end{subequations}
where we moved to conformal time $\tau=\int dt/a(t)$. The variables $\Psi$, $\Phi$, $B$, $E$ denote scalar quantities and $D_{ij}\equiv \partial_{i}\partial_{j}-\frac{1}{3}\delta_{ij}\nabla^{2}$.

We choose the conformal Newtonian gauge, where the only non-zero quantities are the potentials $\Psi$ and $\Phi$. Moreover, we can set $\Phi=\Psi$, since there is no anisotropic stress to linear order in our single-field scenario \cite{bran, PertInf}. This implies that the first-order FRW perturbed line-element is given by
\begin{equation}
ds^{2}=a^{2}\left(\tau\right)\left[\left(1+2\Psi\right)d\tau^{2}-\left(1-2\Psi\right)\delta_{ij}dx^{i}dx^{j}\right].\label{perturbedemetric}
\end{equation}
In conformal time, the equation of motion for fluctuations acquires the form
\begin{equation}
\frac{1}{\sqrt{-g}}\partial_{\mu}\left(\sqrt{-g}g^{\mu\nu}\partial_{\nu} \delta \phi\right)+6\xi \frac{a''}{a^{3}} \delta \phi+\chi (\delta\phi)^{3}-\frac{4\Lambda^{4}}{v^{2}}\delta \phi=0, \label{backeq}
\end{equation}
where we  introduced $\Lambda^{4}=\chi v^{4}/4$. Expanding, as usual, field and metric fluctuations in Fourier modes
\begin{equation}
\delta \phi\left(\textbf{x},\tau\right)=\delta \phi_{k}\left(\tau\right) e^{i\textbf{k}\cdot \textbf{x}}\hspace{2mm},\hspace{2mm} \Psi\left(\textbf{x},\tau\right)=\Psi_{k}\left(\tau\right) e^{i\textbf{k}\cdot \textbf{x}},\label{Ansatz}
\end{equation}
we obtain\footnote{For additional details on the inflaton dynamics in conformal time and linearized perturbed equations, see Appendix~\ref{AppA}.}

\begin{equation}
\begin{split}
&\delta \phi''_{k}+2\mathcal{H}\delta \phi'_{k}+k^{2}\delta \phi_{k}-4\Psi'_{k}\phi'_{k}\\
&=-\xi \bigl(2k^{2}\Psi_{k}-6\Psi''_{k}-24\mathcal{H}\Psi'_{k}-12\frac{a''}{a}\Psi_{k}-4k^{2}\Psi_{k}\bigr)\phi\\
&-\left(V^{\rm eff}_{,\phi \phi}\delta \phi_{k}+2\Psi_{k}V^{\rm eff}_{,\phi}\right)a^{2}.\label{ddotphiNNN}
\end{split}
\end{equation}
Each $k$-mode in Eq. \eqref{Ansatz} evolves independently to leading order.

In this respect, denoting with $H_I$ the Hubble parameter during inflation, one can show that inflaton fluctuations oscillate on sub-Hubble scales $k \gg a(\tau) H_I$, while they freeze out after horizon crossing, on super-Hubble scales $k \ll a(\tau) H_I$ \cite{PertInf}. Hence, fluctuations of cosmological interest today were mainly generated at sub-Hubble scales, but propagated at super-Hubble scales for a long  interval of time. Analogously, we may also expect super-Hubble modes to mostly contribute to  geometric particle production, as we will show later. On super-Hubble scales, the perturbation potential is
\begin{equation}
    \Psi_{k}\simeq \epsilon \mathcal{H} \frac{\delta \phi_{k}}{\phi^\prime},\label{PerPot}
\end{equation}
where $\epsilon=1-\mathcal{H'}/\mathcal{H}^{2}$ is the slow-roll parameter. As discussed above, perturbations on super-Hubble scales are approximately frozen. Moreover, we require $\xi\ll 1$ in order to successfully realize inflation for a potential under the  form\footnote{As we will see, this will specify the self-coupling constant $\chi$ in terms of $\xi$.} of Eq. \eqref{effpot}, in case of positive coupling constant to curvature \cite{ssb3}. At the same time, if the coupling constant is too small, this may reduce significantly the amount of produced particles, as we will see.

Bearing these prescriptions in mind, Eq.~\eqref{ddotphiNNN} becomes
\begin{equation}
\delta \phi''_{k}+2\mathcal{H}\delta \phi'_{k}+\left[k^{2}+\left(V^{\rm eff}_{,\phi\phi}+2\epsilon \frac{\ \mathcal{H}}{\phi'}V^{\rm eff}_{,\phi}\right)a^{2}\right]\delta \phi_{k}=0.\label{newscalarfieldeq}
\end{equation}
Rescaling the field by $\delta\phi_{k}\rightarrow \delta \chi_{k}=\delta \phi_{k}a$,  Eq.~\eqref{newscalarfieldeq} gives
\begin{equation}
\delta \chi''_{k}-\frac{a''}{a}\delta \chi_{k}+\left[k^{2}+\left(V^{\rm eff}_{,\phi\phi}+2\epsilon \frac{\mathcal{H}}{\phi'}V^{\rm eff}_{,\phi}\right)a^{2}\right]\delta \chi_{k}=0,\label{ddotphiNNNNN}
\end{equation}
that explicitly does not depend on the potential offset. To address the cosmological constant problem, however, we delve into its role during and after inflation.

\subsection{Choosing the potential offset } \label{Sez2C}

To manifestly deal with the potential offset, let us first specify the inflationary potential in Eq. \eqref{ddotphiNNNNN}.  To do so, imposing the slow-roll condition
\be \label{slorol}
3\mathcal{H} \phi^\prime \simeq -V^{\rm eff}_{, \phi}a^2,
\ee
we obtain
\begin{equation}
\delta \chi''_{k}+\left[k^{2}+V^{\rm eff}_{,\phi\phi}a^{2}-\frac{a''}{a}-6\epsilon \left(\frac{a'}{a}\right)^{2}\right]\delta \chi_{k}=0.\label{pertbationrescaledeq}
\end{equation}
Substituting now $V^{\rm eff}\left(\phi,R\right)$ in its explicit parts, this gives
\begin{equation}
    \delta \chi''_{k}+\left[k^{2}+V_{,\phi\phi}a^{2}-\left(1-6\xi\right)\frac{a''}{a}-6\epsilon \left(\frac{a'}{a}\right)^{2}\right]\delta \chi_{k}=0,\label{pertbationrescaledeqnew}
\end{equation}
 where we have explicitly written the scalar curvature in conformal time, namely $R=6a^{\prime \prime}/a^3$.
During slow-roll, a suitable choice for the inflationary scale factor is provided by a quasi-de Sitter ansatz\footnote{Further mathematical details are discussed in Appendix \ref{AppB}.}
\be \label{quasiDS}
 a\left(\tau\right)=-\frac{1}{H_{I}}\frac{1}{\tau^{\left(1+\epsilon\right)}},
\ \ \ \ \ \tau < 0,
\ee
where $H_I$ describes the Hubble parameter during inflation and $\epsilon$ quantifies deviations from a pure de Sitter phase. This ansatz approximates an expanding phase dominated by a quasi-constant vacuum energy term, avoiding the direct numerical calculation of the scale factor from the inflaton potential of Eq. \eqref{effpot}. In other words, the parameter $\epsilon$ is required since the slow-roll dynamics of the inflaton field affects the amount of vacuum energy available during inflation, which in turn allows for particle production.

From Eqs. \eqref{pertbationrescaledeqnew}-\eqref{quasiDS} with $\epsilon\ll1$, we obtain
\begin{equation}
 \delta \chi''_{k}+\left[k^{2}-\frac{1}{\tau^{2}}\left(-\frac{V_{,\phi\phi}}{H_{I}^{2}}+\left(1-6\xi\right)\left(2+3\epsilon\right)+6\epsilon\right)\right]\delta \chi_{k}=0.\label{ddotphibas}
\end{equation}
As mentioned earlier, the scalar field carries on vacuum energy and so the presence of the offset term, $V_0$,  appears essential, since it affects the overall vacuum energy during the phase transition, \cite{Martin}. Hence, at this stage, it appears crucial to characterize the role of the offset within the inflationary dynamics.

Obviously, if we assume $V_0=0$, both large and small $\phi$ are allowed during slow-roll \cite{ssb2}. Additionally, as shown by Eq. \eqref{Latot}, a negligible offset also implies that after inflation we recover the bare contribution to the cosmological constant, namely $\Lambda_B$.  However, if we assume the existence of a cosmological constant related to vacuum energy, intuitively we expect that a small offset may have a similar magnitude than $\phi$. In summary, we can identify two main characteristics of the offset term $V_0$:

\begin{itemize}
    \item[-] if the offset is $V_{0}=0$, there is a constant term in the inflaton potential that, if  sufficiently large, can be interpreted as vacuum energy \emph{before the transition} and reads $\Lambda^4= \chi v^4/4$. In this case, it drives inflation and also determines the scalar curvature, in fact $R=8\pi G\left(4 \Lambda^4+T^{\mu(\phi)}_{\mu}\right)$, where $T^{\mu(\phi)}_\mu$ is the trace of the energy-momentum tensor for the scalar field  during the  slow-roll. Here, we require $\Lambda^{4}\sim 10^{64}$ $\unit{\giga\electronvolt}^{4}$ in order to have a suitable vacuum energy contribution to drive inflation, thus working out a \emph{small-field scenario};
    \item[-] if $V_0 \neq 0$, the situation is very different. Setting $V_{0}\simeq -\Lambda^{4}$ with the aim of cancelling vacuum energy, inflation can be realized only in a \emph{large-field} scenario. Accordingly, inflation is driven directly by the inflaton that transports the energy $\chi\phi^{4}/4$. Consequently, vacuum energy is reinterpreted as field-dependent, say $\Lambda^{4}\left(\phi\right)$. Thus, applying this choice for the potential offset, vacuum energy is not a constant term during inflation and the equation of state for the field reads $P^{\phi}/\rho^{\phi}\neq-1$, violating the no-go theorem \cite{nogo} during the phase transition, which is reinterpreted as a metastable phase.
\end{itemize}
In summary, both small and large fields are in principle viable to describe a symmetry breaking mechanism that transports vacuum energy, thought as responsible for the inflationary phase.

Based on these reasons, we proceed to characterize inflation in the context of both small and large inflaton fields, also incorporating the concept of geometric particle production. We then explore the implications of generating dark matter constituents within these two scenarios, considering the two above options for the offset and with the primary objective to achieve vacuum energy cancellation at the conclusion of the slow-roll phase.


\section{Small-field symmetry breaking inflation} \label{sezione3}

Within the small-field scenario, the potential minimum is placed at $\phi=0$ before the phase transition.
Hence, in the slow-roll, having $f(\epsilon,\xi)\equiv \left(1-6\xi\right)\left(2+3\epsilon\right)+6\epsilon$,  the field is close to zero and
Eq.~\eqref{ddotphibas} becomes
\begin{equation}
\begin{split}
    \delta \chi''_{k}&+\biggl[k^{2}-\frac{1}{\tau^{2}}\biggl(f(\epsilon,\xi)+\frac{4 \Lambda^{4}}{v^{2}H_{I}^{2}}-\frac{3\chi\phi^{2}}{H_{I}^{2}}\biggr)\biggr]\delta \chi_{k}=0.\label{smallfieldpert}
\end{split}
\end{equation}
Here $\phi \ll v$ therefore we can safely neglect the last quadratic contribution, to have
\begin{equation}
    \delta \chi''_{k}+\left[k^{2}-\frac{1}{\tau^{2}}\left(\nu^{2}-\frac{1}{4}\right)\right]\delta \chi_{k}=0,\label{Pertscaledequation}
\end{equation}
where $\nu^{2}$ explicitly reads
\begin{equation}
\nu^{2}=\frac{1}{4}+\frac{4\Lambda^{4}}{v^{2}H_{I}^{2}}+f(\epsilon,\xi).\label{coeffnuour}
\end{equation}
The standard solution of  Eq.~\eqref{Pertscaledequation} is
\begin{equation}
    \delta \chi_{k}\left(\tau\right)=\sqrt{-\tau}\left[c_{1}(k)H^{\left(1\right)}_{\nu}\left(-k\tau\right)+c_{2}(k)H^{\left(2\right)}_{\nu}\left(-k\tau\right)\right],\label{pertequres}
\end{equation}
in which $c_{1}(k)$, $c_{2}(k)$ are integration constants, while $H^{(1)}_{\nu}$, $H^{(2)}_{\nu}$ are \emph{Hankel functions} of the first and second kind, respectively.

The constants $c_1(k)$ and $c_2(k)$ are specified by selecting the initial vacuum state of the field. Among the various possibilities \cite{vacchoice}, a common choice is the Bunch-Davies vacuum, which represents a local attractor in the space of initial states for an expanding background \cite{bran, Bunchvac0, Bunchvac1, Bunchvac2}. It implies that, in the ultraviolet regime $k \gg a H_I$, the inflaton fluctuation $\delta \chi_k$ matches the plane-wave solution
\begin{equation}
    \delta \chi_{k}\sim \frac{e^{-ik\tau}}{\sqrt{2k}}.
\end{equation}
Thus, considering the asymptotic Hankel functions,
\begin{align}
H^{\left(1\right)}_{\nu}\left(x\gg1\right)&\simeq\sqrt{\frac{2}{\pi x}}e^{i\left(x-\frac{\pi}{2}\nu-\frac{\pi}{4}\right)},\\
H^{\left(2\right)}_{\nu}\left(x\gg1\right)&\simeq\sqrt{\frac{2}{\pi x}}e^{-i\left(x-\frac{\pi}{2}\nu-\frac{\pi}{4}\right)},
\end{align}
we can take the ultraviolet limit $-k\tau\gg1$ of Eq. \eqref{pertequres} to derive the initial conditions for fluctuations, which read \cite{PertInf}
\begin{subequations}\label{coeffck}
   \begin{align}
       &c_{1}(k)=\frac{\sqrt{\pi}}{2}e^{i\left(\nu+\frac{1}{2}\right)\frac{\pi}{2}},\\
       &c_{2}(k)=0.
    \end{align}
\end{subequations}

\subsection{Super-Hubble scales} \label{Sez3A}

We now focus on super-Hubble fluctuations. As anticipated in Sec. \ref{Sez2B}, a ``classical" description of inflaton fluctuations becomes valid after horizon crossing, where the field modes cease to oscillate in time. Accordingly, particle production should be enhanced on super-Hubble scales with respect to sub-Hubble ones, since in this latter case the oscillating nature of fluctuations is expected to reduce probability amplitudes, as we will see. The same enhancement was recently obtained in the case of gravitational particle production from vacuum \cite{boya}. We also remark that super-Hubble modes are typically less sensitive to the background dynamics at the end of inflation. In fact, wavelengths larger than the particle horizon are causally disconnected from the microphysical processes of thermalization that should take place during reheating \cite{reh1, reh2}.

On super-Hubble scales, the Hankel function entering fluctuations is given by
\begin{equation}
H^{\left(1\right)}_{\nu}\left(x\ll1\right)\simeq \sqrt{\frac{2}{\pi}}e^{-i\frac{\pi}{2}}2^{\left(\nu-\frac{3}{2}\right)}\frac{\Gamma\left(\nu\right)}{\Gamma\left(\frac{3}{2}\right)}x^{-\nu},
\end{equation}
and thus we obtain
\begin{equation}
    \delta \chi_{k}=e^{i\left(\nu-\frac{1}{2}\right)\frac{\pi}{2}}2^{\left(\nu-\frac{3}{2}\right)}\frac{\Gamma\left(\nu\right)}{\Gamma\left(\frac{3}{2}\right)}\frac{1}{\sqrt{2k}}\left(-k\tau\right)^{\frac{1}{2}-\nu}.\label{fluctuationssolution}
\end{equation}
Restoring then the original perturbation $\delta \phi_{k}$, we finally have
\begin{equation}
    \delta \phi_{k}=e^{i\left(\nu-\frac{1}{2}\right)\frac{\pi}{2}}2^{\left(\nu-\frac{3}{2}\right)}\frac{\Gamma\left(\nu\right)}{\Gamma\left(\frac{3}{2}\right)}\frac{H_{I}}{\sqrt{2k^{3}}}\left(\frac{k}{aH_{I}}\right)^{\frac{3}{2}-\nu}.\label{Perphi}
\end{equation}
Since $\epsilon, \xi \ll 1$, from Eq. \eqref{coeffnuour} we observe that $\nu \simeq 3/2$ for the typical energy scales of inflation. This implies that inflation fluctuations are almost frozen in on super-Hubble scales, acquiring only a tiny dependence on time after horizon crossing. Now, in order to determine the geometric perturbation of Eq.~\eqref{PerPot} on these scales, we have to solve also the background field equation.

Exploiting the slow-roll condition of Eq. \eqref{slorol}, we can write
\begin{equation}
    3\left(\frac{1+\epsilon}{\tau}\right)\phi^\prime=-\left(\chi\phi^{3}-\frac{4\Lambda^{4}}{v^{2}}\phi+6\xi \frac{a''}{a^{3}}\phi\right)a^{2},
\end{equation}
and using the quasi-de Sitter scale factor, we find
\begin{equation}
3\left(1+\epsilon\right)\phi'=-\chi\phi^{3}\frac{1}{H_{I}^{2}\tau}-\frac{4\Lambda^{4}}{v^{2}}\frac{\phi}{H_{I}^{2}\tau}+6\xi \left(2+3\epsilon\right)\frac{\phi}{\tau}.
\end{equation}

\begin{figure}[H]
\begin{center}
\includegraphics[width=8cm]{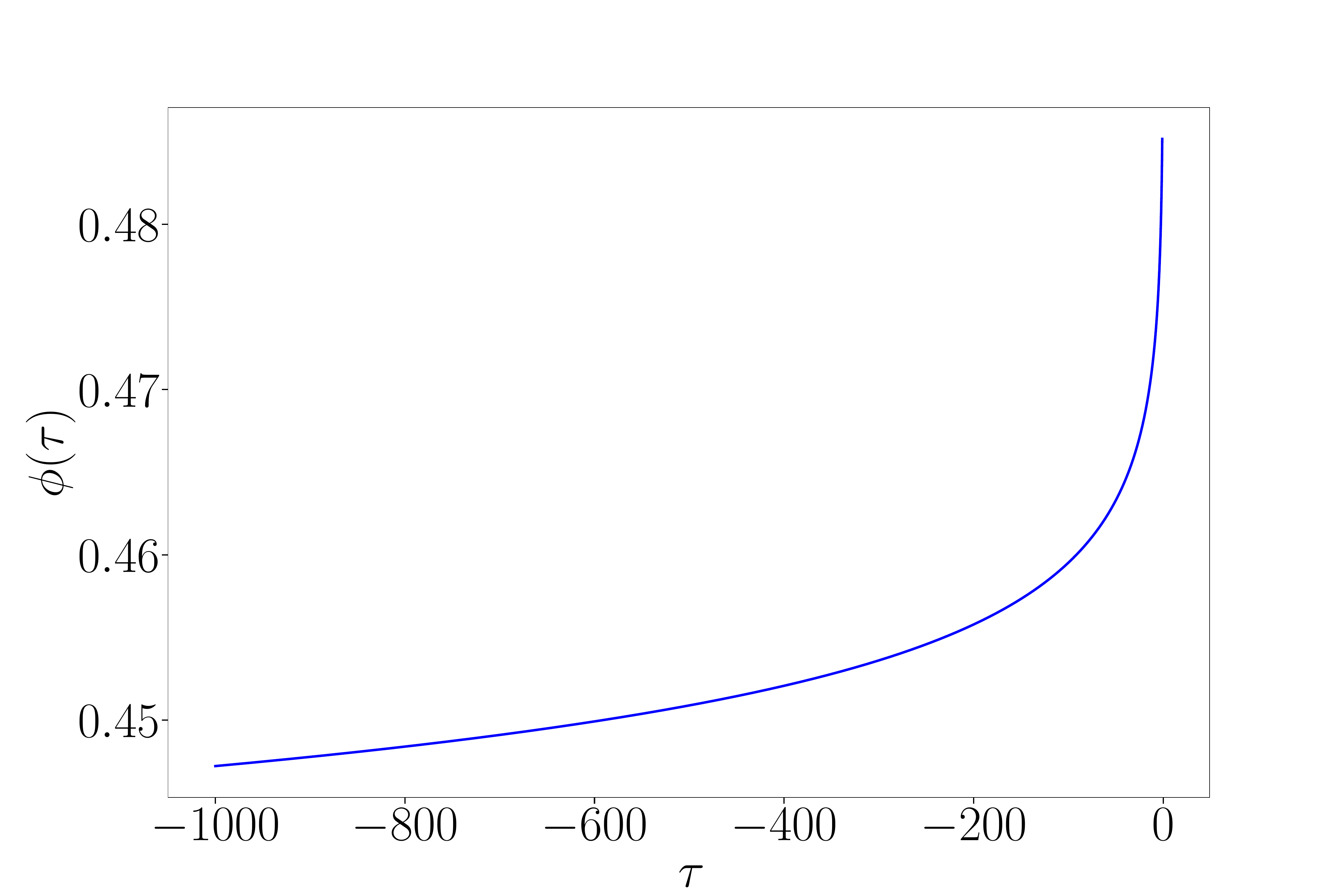}
\caption{Background field evolution with $\Lambda^{4}=10^{64}$ \unit{\giga\electronvolt}$^{4}$ and $\chi=10^{-14}$. The field $\phi\left(\tau\right)$ is chosen such that $\phi\left(\tau_{i}\right)=v/10^{20}$, with $\tau_i=10^{3}$ GeV$^{-1}$. The other parameters are: $\epsilon=10^{-3}$ and $\xi=10^{-5}$. The values of $\chi$ and $\xi$ are chosen so that the Newtonian constant is not significantly screened by non-minimal coupling.}
\label{fielback}
\end{center}
\end{figure}

As above stated, the field value during slow-roll is close to zero. Accordingly, we neglect terms proportional to $\sim\phi^3$, to obtain
\begin{equation}
\phi\left(\tau\right)=c_{0}\left|\tau\right|^{\frac{-\eta +2\xi\left(2+3\epsilon\right)}{1+\epsilon}}=c_{0}\left|\tau\right|^{j},
\end{equation}
where in the last expression we have introduced $\eta \equiv4\Lambda^{4}/3H_{I}^{2}v^{2}$ and
$j\equiv \left[-\eta+2\xi\left(2+3\epsilon\right)\right]/(1+\epsilon)$. The parameter $c_0$ is determined by selecting the initial value for the background field.

In Fig. \ref{fielback} we show the slow-roll dynamics of the inflaton, for the initial condition $\phi(\tau_i)=v/10^{20}$ and $\tau_i=10^{3}$ GeV$^{-1}$.

\begin{figure}[H]
\begin{center}
\includegraphics[width=8cm]{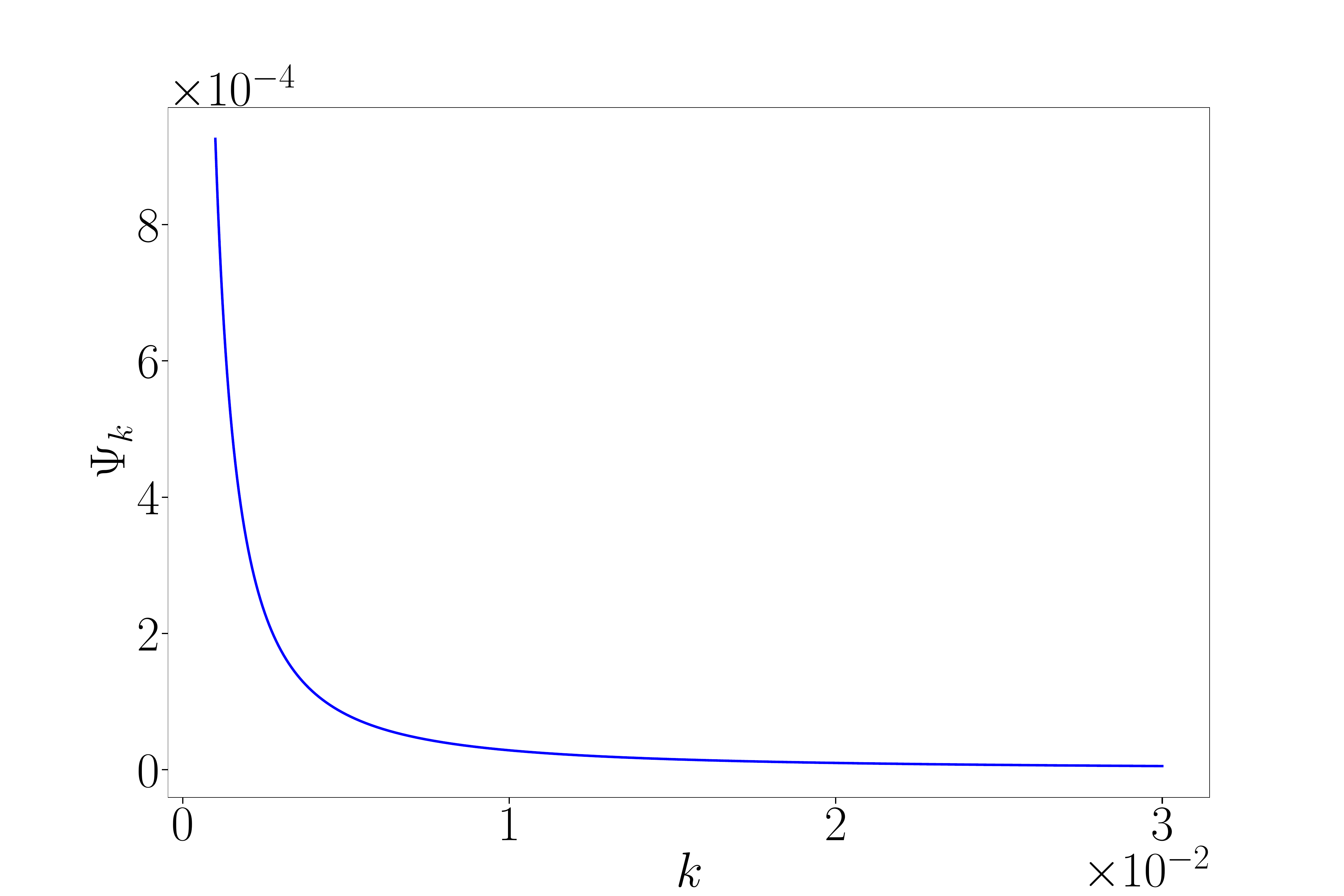}
\caption{Perturbation potential $\Psi_k$ during slow-roll, with $\Lambda^{4}=10^{64}$ \unit{\giga\electronvolt}$^{4}$ and $\chi=10^{-14}$. The other parameters are: $\xi=10^{-5}$, $\epsilon=10^{-3}$ and $\alpha=10^{20}$. The value of the parameter $\alpha$ is chosen so that the perturbation is normalized in analogy to the field fluctuations. }\label{zeta}
\end{center}
\end{figure}

Similarly, in order to have fluctuations comparable to the background, we need to rescale the coefficients $c(k)$ in Eqs.~\eqref{coeffck} so that
\begin{equation}
    \delta \phi \rightarrow \frac{1}{\alpha}\delta \phi.
\end{equation}
Such normalization is required to satisfy the condition $\left| \delta \phi \right| \ll \phi$ throughout the slow--roll phase. This readily gives
\begin{equation}
\begin{split} \label{pertuSH}
    \Psi_{k}\left(\tau\right)=-\frac{\epsilon}{\alpha}e^{i\left(\nu-\frac{1}{2}\right)\frac{\pi}{2}}2^{\nu-\frac{3}{2}}\frac{\Gamma\left(\nu\right)}{\Gamma\left(3/2\right)}\frac{\mathcal{H}}{\left(jc_{0}\left|\tau\right|^{j-1}\right)}\times
    \\\frac{H_{I}}{\sqrt{2k^{3}}}\left(\frac{k}{aH_{I}}\right)^{\frac{3}{2}-\nu},
\end{split}
\end{equation}
which is plotted in Fig.~\ref{zeta}.

Selecting now a specific number of e-foldings, $N\geq 60$, that agrees with current most-accepted values \cite{Planck}, we can derive the time $\tau_f$ at which inflation is expected to end, by
\begin{equation}\label{efoldings}
    N=\int dt H(t)\simeq -\int ^{\tau_{f}}_{\tau_{i}} d\tau \frac{H_{I}}{H_{I}\tau}=60.
\end{equation}
Since we primarily focus on super-Hubble scales, we need to introduce a cut-off time $\tau^\prime$ at which modes $k< a(\tau^\prime) H_I$ already crossed the horizon and then study the evolution of these modes until the end of inflation, i.e., in the interval $[\tau^\prime, \tau_f]$.

Geometric particle production due to inflaton fluctuations can be quantified resorting to a perturbative approach \cite{fri, ces}. We start from the first-order Lagrangian describing interaction between fluctuations and spacetime perturbations,
\begin{equation}
    \mathcal{L}_{I}=-\frac{1}{2}\sqrt{-g_{(0)}}H^{\mu\nu}T^{\left(0\right)}_{\mu\nu}, \label{intlag}
\end{equation}
where $H^{\mu \nu}$ is related to the metric perturbations and $T_{\mu \nu}^{(0)}$ is the zero-order energy-momentum tensor for the field fluctuations\footnote{All the mathematical details concerning geometric particle production are reported in Appendix \ref{AppC}.}. By expanding the corresponding $\hat S$ matrix at first order in Dyson series, one can show that the number density of particles computed at a specific time $\overline{\tau}$ is
\begin{equation}
    N^{(2)}\left(\overline{\tau}\right)=\frac{1}{\left(2\pi a\left(\overline{\tau}\right)\right)^{3}}\int d^{3}q d^{3}p \left|\braket{0|\hat{S}|p,q}\right|^{2},\label{GNP}
\end{equation}
where normalization is over the comoving volume and  $\langle 0 \lvert \hat{S} \rvert p, q\rangle$ is the first-order probability amplitude associated to particle pair production with momenta $p$ and $q$. In Eq. \eqref{GNP}, we have neglected the contribution due to nonperturbative gravitational particle production, as discussed later on (see also Appendix \ref{AppC}).

In the small-field regime, the symmetry breaking potential is approximately given by
\begin{equation}
V\left(\phi\right)\simeq\Lambda^{4}-\frac{2\Lambda^{4}}{v^{2}}\phi^{2},
\end{equation}
thus the zero-order energy-momentum tensor associated to fluctuations is indicated as follows
\begin{align}
T^{\left(0\right)}_{\mu\nu}\simeq
&\partial_{\mu}\delta\phi\partial_{\nu}\delta\phi\notag \\&-\frac{1}{2}g_{\mu\nu}^{\left(0\right)}\bigg[g^{\rho\sigma}_{\left(0\right)}\partial_{\rho}\delta\phi\partial_{\sigma}\delta\phi
-2\Lambda^{4}+\frac{4\Lambda^{4}}{v^{2}}\left(\delta\phi\right)^{2}\bigg] \notag \\
&-\xi \left[\nabla_{\mu}\partial_{\nu}-g_{\mu\nu}^{\left(0\right)}\nabla^{\rho}\nabla_{\rho}+R^{\left(0\right)}_{\mu\nu}-\frac{1}{2}R^{\left(0\right)}g_{\mu\nu}^{\left(0\right)}\right]\left(\delta \phi\right)^{2}.
\end{align}
Accordingly, the total probability amplitude for pair production is
\begin{equation}
\begin{split}
    &\braket{p,q|\hat{S}|0}=-\frac{i}{2}\int d^{4}x 2a^{4}H^{\mu\nu}\biggl[\partial\left(_{\mu}\delta\phi^{*}_{p}\delta_{\nu}\right)\delta\phi^{*}_{q}\\
    &-\frac{1}{2}\eta_{\mu\nu}\eta^{\rho\sigma}\partial\left(_{\rho}\delta\phi^{*}_{p}\partial_{\sigma}\right)\delta \phi^{*}_{q}-g_{\mu\nu}^{\left(0\right)}\frac{2\Lambda^{4}}{v^{2}}\delta\phi^{*}_{p}\delta \phi^{*}_{q}\\
    &-\xi\left(\nabla_{\mu}\partial_{\nu}-g_{\mu\nu}^{\left(0\right)}\nabla^{\rho}\nabla_{\rho}+R^{\left(0\right)}_{\mu\nu}-\frac{1}{2}R^{\left(0\right)}g^{\left(0\right)}_{\mu\nu}\right)\delta\phi^{*}_{p}\delta\phi^{*}_{q}\biggr]\\
    &\times e^{-i\left(\textbf{p}+\textbf{q}\right)\cdot \textbf{x}}\label{probampl},
\end{split}
\end{equation}
which requires proper normalization with respect to the zero-order matrix element for field fluctuations.

We also remark that time integration has to be performed in the interval $[\tau^\prime, \tau_f]$ previously discussed. This implies that modes of interest are inside the Hubble horizon at the beginning of inflation, but they are all in super-Hubble form starting from $\tau=\tau^\prime$, namely\footnote{This approach inevitably leads to underestimate the total number of particles produced, since we neglect the contribution of modes that exit Hubble horizon after the cut-off time $\tau^\prime$. In Appendix \ref{AppD}, we discuss more in detail how super-Hubble production is influenced by the choice of $\tau^\prime$. We plan to come back to this point in future works, in order to refine our estimate of geometric particles produced during slow-roll.}
\be
\label{momcond}
a(\tau_i) H_I < k < a(\tau^\prime) H_I.
\ee

Thus, Eq. \eqref{probampl} can be written more compactly as
\begin{align} \label{compact}
\langle p, q \lvert \hat{S} \rvert 0 \rangle=-\frac{i}{2}\int d^{4}x 2a^{2}\bigg(&A_{0}\left(\textbf{x},\tau\right)+A_{1}\left(\textbf{x},\tau\right) \notag \\
&+A_{2}\left(\textbf{x},\tau\right)+A_{3}\left(\textbf{x},\tau\right)\bigg),
\end{align}
where
\begin{equation}
\begin{split}   &A_{0}\left(\textbf{x},\tau\right)=2\Psi\biggl[\partial_{0}\delta\phi^{*}_{p}\partial_{0}\delta\phi^{*}_{q}-\frac{1}{2}\eta^{\rho\sigma}\partial_{\rho}\delta\phi^{*}_{p}\partial_{\sigma}\delta\phi^{*}_{q}\\
&-2a^{2}\frac{\Lambda^{4}}{v^{2}}\delta\phi^{*}_{p}\delta\phi^{*}_{q}-\xi\biggl(\partial_{0}\partial_{0}-\frac{a'}{a}\partial_{0}-\eta^{\rho\sigma}\partial_{\rho}\partial_{\sigma}\\
&-3\left(\frac{a'}{a}\right)^{2}\biggr)\delta\phi^{*}_{p}\delta\phi^{*}_{q}\biggr] e^{-i\left(\textbf{p}+\textbf{q}\right)\cdot \textbf{x}},\label{A0}
\end{split}
\end{equation}
and
\begin{equation}
\begin{split}   &A_{i}\left(\textbf{x},\tau\right)=2\Psi\biggl[\partial_{i}\delta\phi^{*}_{p}\partial_{i}\delta\phi^{*}_{q}+\frac{1}{2}\eta^{\rho\sigma}\partial_{\rho}\delta\phi^{*}_{p}\partial_{\sigma}\delta\phi^{*}_{q}\\
&+2a^{2}\frac{\Lambda^{4}}{v^{2}}\delta\phi^{*}_{p}\delta\phi^{*}_{q}-\xi\biggl(\partial_{i}\partial_{i}+3\frac{a'}{a}\partial_{0}+\eta^{\rho\sigma}\partial_{\rho}\partial_{\sigma}\\
&+2\frac{a''}{a}-\left(\frac{a'}{a}\right)^{2}\biggr)\delta\phi^{*}_{p}\delta\phi^{*}_{q}\biggr] e^{-i\left(\textbf{p}+\textbf{q}\right)\cdot \textbf{x}},\label{Ai}
\end{split}
\end{equation}
for $i=1,2,3$.

Finally, in order to obtain numerical values for the total number of particles, we need constraints over vacuum energy density, $\Lambda^{4}$, and self-coupling constant, $\chi$.

On the one hand, if we assume vacuum energy domination during inflation, the Hubble rate reads
\begin{equation}
    H^{2}(t)\equiv H_{I}^{2}\simeq \frac{8\pi G}{3}\Lambda^{4},\label{HIinflation}
\end{equation}
and, following the Planck satellite results \cite{Planck}, we fix
\begin{equation}
    \frac{H_{I}}{M_{P}}\simeq 2.5 \cdot 10^{-5},\label{Plancksatellite}
\end{equation}
so that vacuum energy needs to satisfy $\Lambda^4\leq 10^{65}$ \unit{\giga\electronvolt^4}, in order to avoid fine-tuning.

Furthermore, the self-coupling constant, $\chi$, is usually constrained as function of $\xi$ in order to ensure density inhomogeneities of the proper size during inflation \cite{ssb2}. One usually requires a very small ratio
\be \label{constra}
\sqrt{\chi/\xi^2} \sim 10^{-5},
\ee
to have agreement with recent observational data \cite{Nonmincoup0}. However, for small-field inflation this ratio cannot be fulfilled without drastically altering the value of the gravitational constant $G$.

For example, selecting $\Lambda^{4}=10^{64}$ \unit{\giga\electronvolt}$^{4}$ and asking $v$ to be of the order of Planck mass, namely $v^2 \sim 10^{39}$ GeV$^2$, we would have $\chi \sim 10^{-14}$. Then, only assuming a very small coupling constant $\xi \leq 10^{-5}$, we get
\begin{equation}
    8\pi G \xi v^{2} \leq 10^{-3},
\end{equation}
so that field-curvature coupling would not significantly screen Newton's gravitational constant after inflation. This choice, however, leads to a significant violation of the condition in Eq. \eqref{constra}.

\begin{figure}[H]
        \includegraphics[width=8cm]{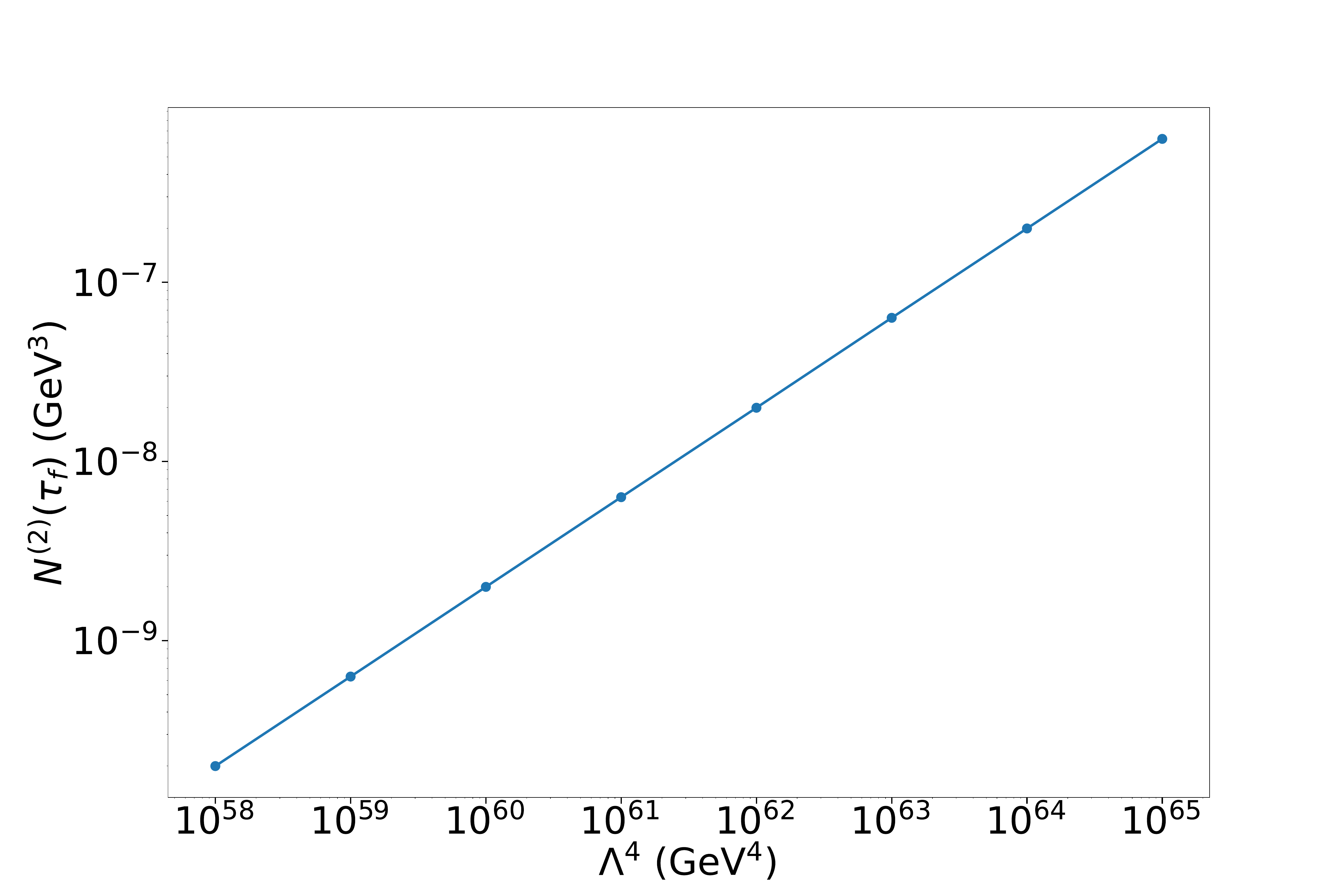}
    \caption{Number density $N^{(2)}$ in $\unit{\giga\electronvolt}^{3}$ as function of vacuum energy $\Lambda^{4}$ in loglog scale. The number density is computed assuming $N=60$ e-foldings, $\tau_{i}=-10^{3}$ \unit{\giga\electronvolt}$^{-1}$, $\tau'=-5\times 10^{-2}$ \unit{\giga\electronvolt}$^{-1}$,  $\epsilon=10^{-3}$ and $\xi=10^{-5}$. }
    \label{VarLsmall}
\end{figure}

Fig.~\ref{VarLsmall} shows the number density of particles produced during inflation as function of vacuum energy $\Lambda^{4}$, for $\tau'=-5\times 10^{-2}$ \unit{\giga\electronvolt}$^{-1}$.
The values of $\Lambda^{4}$, $\chi$ and the corresponding number densities are collected in Tab.~\ref{Tab1}.

\begin{table}[H]
\begin{center}
\begin{tabular}{|| c | c | c||}
\hline
 $\Lambda^{4}$ (\unit{\giga\electronvolt})$^{4}$  & $\chi$ & $N^{(2)}\left(\tau_{f}\right)$
 (\unit{\giga\electronvolt})$^{3}$\\ [0.5ex]
 \hline\hline
 $\num{1.0}\times 10^{58}$ & $1.0\times 10^{-20}$ & $\num{2.0}\times 10^{-10}$  \\
 $\num{1.0}\times 10^{59}$& $1.0\times 10^{-19}$ &  $\num{6.3}\times 10^{-10}$ \\
 $\num{1.0}\times 10^{60}$ & $1.0\times 10^{-18}$ & $\num{2.0}\times 10^{-9}$ \\
 $\num{1.0}\times 10^{61}$ & $1.0\times 10^{-17}$ & $\num{6.3}\times 10^{-9}$ \\
 $\num{1.0}\times 10^{62}$ & $1.0\times 10^{-16}$ & $\num{2.0}\times 10^{-8}$ \\
 $\num{1.0}\times 10^{63}$ & $1.0\times 10^{-15}$ & $\num{6.3}\times 10^{-8}$ \\
 $\num{1.0}\times 10^{64}$   & $1.0\times 10^{-14}$ & $\num{2.0}\times 10^{-7}$ \\
 $\num{1.0}\times 10^{65}$ & $1.0\times 10^{-13}$ & $\num{6.3}\times 10^{-7}$ \\
\hline
\end{tabular}
\end{center}
\caption{Number density of geometric particles produced in the small-field scenario, for different values of vacuum energy $\Lambda^4$ and corresponding self-coupling constant $\chi$.}
\label{Tab1}
\end{table}

\section{Large-field symmetry breaking inflation} \label{sezione4}

As above stated, in the large-field inflation, considering the offset $V_{0}=-\Lambda^{4}$, the role of vacuum energy driving inflation is directly played by the self-interacting term $\chi \phi^{4}/4$, thus implying that the field initial value lies around Planck mass.

Within this framework, we can set $\phi\gg v$ during slow-roll \cite{ssb2}, so the potential in Eq.~\eqref{effpot} simplifies to
\begin{equation} \label{potlarg}
    V^{\rm eff}\left(\phi,R\right)\simeq \frac{\chi}{4}\phi^{4}+\frac{1}{2}\xi R\phi^{2}.
\end{equation}

Hence, the background field equation of Eq. \eqref{slorol} now gives
\begin{equation}
  3\mathcal{H}\phi'=-\left(\chi \phi^{3}+\xi R \phi\right)a^{2},\label{Largebackfield}
\end{equation}
and by plugging above the quasi-de Sitter scale factor, Eq. \eqref{quasiDS}, we get
\begin{equation}
3\left(\frac{1+\epsilon}{\tau}\right)\phi'=\frac{\chi}{H_{I}^{2}\tau^{2}}\phi^{3}+6\xi \frac{\left(2+3\epsilon\right)}{\tau^{2}}\phi,
\end{equation}
namely
\begin{equation}
    \phi'-\frac{2\xi \left(2+3\epsilon\right)}{\left(1+\epsilon\right)\tau}\phi=\frac{\chi}{3\left(1+\epsilon\right)H_{I}^{2}\tau}\phi^{3}.\label{Bernoulli}
\end{equation}
Eq.~\eqref{Bernoulli} reads under the form of a \emph{Bernoulli differential equation}, $
    \phi'+p(\tau)\phi=q(\tau)\phi^{n}$, where $p(\tau)=-\frac{2\xi \left(2+3\epsilon\right)}{\left(1+\epsilon\right)\tau}$, $q(\tau)=\frac{\chi}{3\left(1+\epsilon\right)H_{I}^{2}\tau}$ and $n=3$.

\begin{figure}[H]
\begin{center}
\includegraphics[width=8cm]{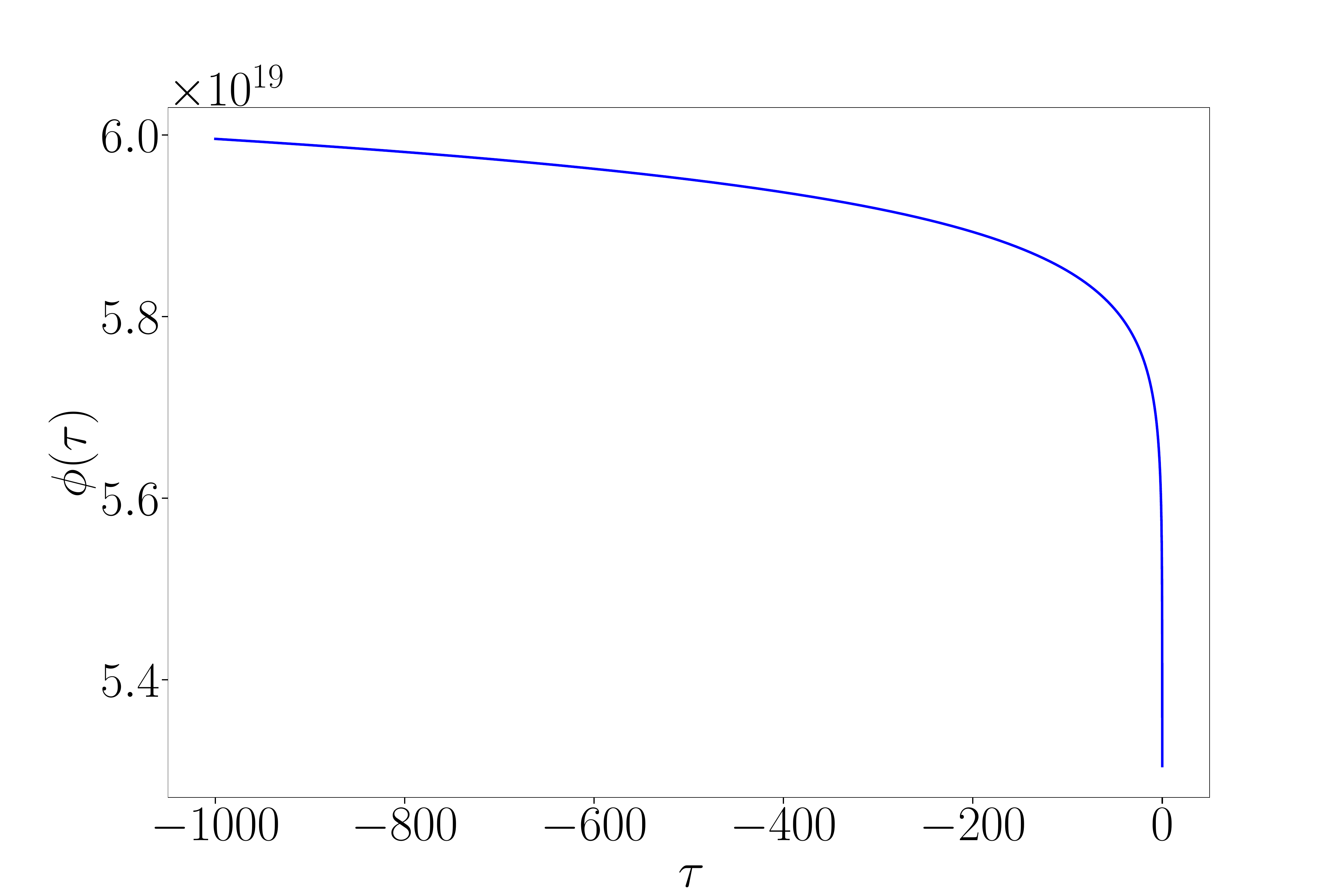}
\caption{Background field evolution with
Hubble rate $H_{I}=4.2\times 10^{12}$ \unit{\giga\electronvolt} and self-coupling constant $\chi=10^{-16}$. The field $\phi\left(\tau\right)$ is chosen by imposing $\phi\left(\tau_{i}\right) = 5\  M_{P}$ as the initial condition for chaotic inflation, with $\tau_i=-10^{3}$ \unit{\giga\electronvolt}$^{-1}$. The other parameters are: $\epsilon=10^{-3}$ and $\xi=10^{-3}$. The values of $\chi$ and $\xi$ are chosen so to respect the constraint for non-minimal coupling inflation, namely Eq. \eqref{constra}.}\label{Largefielback}
\end{center}
\end{figure}

Thus, replacing $\omega=\phi^{1-n}$, we find that  Eq.~\eqref{Bernoulli} becomes a linear differential equation of the form $\frac{1}{1-n}\omega'+p(\tau)\omega=q(\tau)$, reading
\begin{equation}
    \frac{\omega'}{2}+\frac{2\xi\left(2+3\epsilon\right)}{\left(1+\epsilon\right)\tau}\omega=-\frac{\chi}{3\left(1+\epsilon\right)H_{I}^{2}\tau}.\label{Background3}
\end{equation}
By virtue of our comments on large-field inflation and following Ref. \cite{baurev}, we require super-Planckian field values during slow-roll and, so, in Fig.~\ref{Largefielback}, we display the background field, $\phi$, with an indicative initial condition placed around  $\phi\left(\tau_{i}\right)= 5\  M_{P}$.

After identifying the relevant background information, we proceed to analyze the variations. Once again,  Eq.~\eqref{ddotphibas} remains valid when considering the fluctuations of the inflaton.

During slow-roll, both the field and its time derivative are quasi-constant: accordingly, as a plausible treatment to solve Eq.~\eqref{ddotphibas} in the large-field case, we can replace $\phi^{2}$ with its mean value during inflation,
\begin{equation}
    \phi^{2}\rightarrow \frac{\int ^{\tau_{f}}_{\tau_{i}}\phi^{2}d\tau}{\tau_{f}-\tau_{i}}=\braket{\phi^{2}},\label{meanback}
\end{equation}
where the final time $\tau_{f}$, can be derived from $\tau_i$ by selecting the total number of e-foldings, see Eq.~\eqref{efoldings}.

Hence, we obtain
\begin{equation}
\delta \chi''_{k}+\left[k^{2}-\frac{1}{\tau^{2}}\left(f(\epsilon,\xi)-\frac{3\chi \braket{\phi^{2}}}{H_{I}^{2}}\right)\right]\delta \chi_{k}=0,\label{largefieldfluc1}
\end{equation}
and the usual solution, Eq.~\eqref{fluctuationssolution}, is recovered with
\begin{equation}
    \nu^2=\frac{1}{4}-\frac{3\chi \braket{\phi^{2}}}{H_{I}^{2}}+f(\epsilon,\xi).
\end{equation}

\subsection{Super-Hubble scales}

\begin{figure}[H]
\begin{center}
\includegraphics[width=8cm]{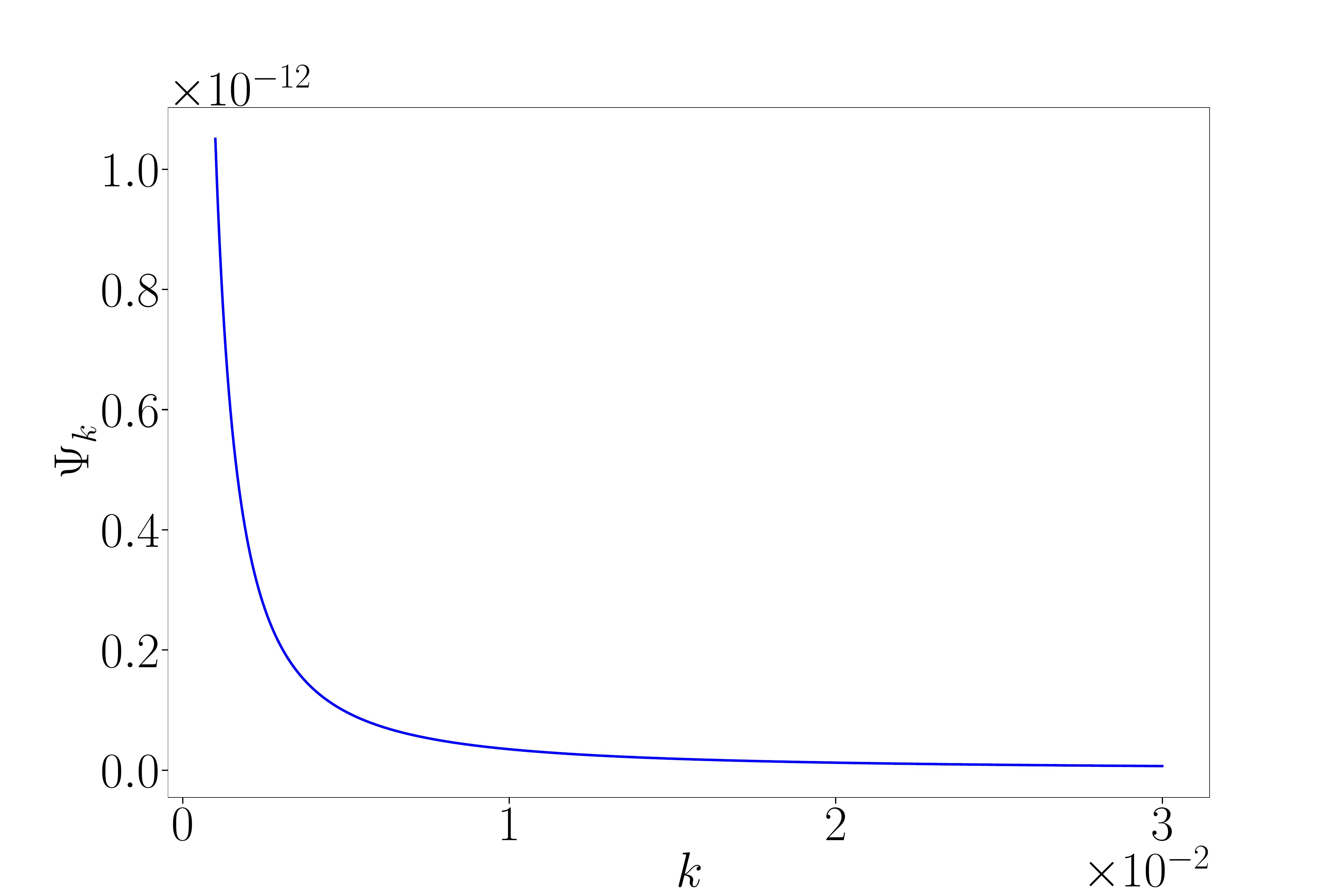}
\caption{Perturbation potential $\Psi_k$ as function of the momentum $k$, with Hubble rate $H_{I}=4.2\times 10^{12}$ \unit{\giga\electronvolt} and self-coupling constant $\chi=10^{-16}$. The other parameters are: $\xi=10^{-3}$ and $\epsilon=10^{-3}$.}\label{Largezeta}
\end{center}
\end{figure}

The geometric perturbation on super-Hubble scales can be expressed again in the form reported in Eq. \eqref{pertuSH}, without the normalizing factor $\alpha$. Its behavior is shown in Fig.~\ref{Largezeta}.

It is worth noticing how to rewrite the (approximate) Lagrangian for large-field, say $\mathcal L_{LF}$, that reads
\begin{equation}
    \mathcal{L}_{LF}\simeq\frac{1}{2}g^{\mu\nu}\partial_{\mu}\phi\partial_{\nu}\phi-\frac{\chi}{4}\phi^{4}-\frac{1}{2}\xi R\phi^{2},\label{LagrLargefield}
\end{equation}
and the zero-order energy-momentum tensor for the fluctuations,
\begin{equation}
\begin{split}    T^{\left(0\right)}_{\mu\nu}&=\partial_{\mu}\delta\phi\partial_{\nu}\delta\phi-\frac{1}{2}g_{\mu\nu}^{\left(0\right)}\left[g^{\rho\sigma}_{\left(0\right)}\partial_{\rho}\delta\phi\partial_{\sigma}\delta\phi-\frac{\chi}{2}\left(\delta\phi\right)^{4}\right]\\
&-\xi \left[\nabla_{\mu}\partial_{\nu}-g_{\mu\nu}^{\left(0\right)}\nabla^{\rho}\nabla_{\rho}+R^{\left(0\right)}_{\mu\nu}-\frac{1}{2}R^{\left(0\right)}g_{\mu\nu}^{\left(0\right)}\right]\left(\delta \phi\right)^{2}.\label{EMTLarge}
\end{split}
\end{equation}
Immediately, from the above Lagrangian and from $T_{\mu\nu}^{(0)}$, we can notice that the potential offset $V_0$ does not enter in the interacting potential. This means that it does not contribute to the amount of geometric particles produced. The same reasoning would apply to the small-field scenario.

Moreover, to single out the most prominent contribution to particle creation, we can naively neglect the quartic term in fluctuations. This counterpart leads to divergences when computing probability amplitudes and thus would require a renormalization procedure. This fact has been severely investigated in a  detailed study of the $\chi \phi^4$ theory in curved spacetime as one can see in Ref. \cite{Birdav}, where different approaches are discussed, including the here-employed interaction picture. In this way, the main contribution to Eq.~\eqref{EMTLarge} is provided by field-curvature coupling and, so, the total probability amplitude for pair production can be written again in the form of Eq. \eqref{compact}, where now
\begin{equation}
\begin{split}   &A_{0}\left(\textbf{x},\tau\right)=2\Psi\biggl[\partial_{0}\delta\phi^{*}_{p}\partial_{0}\delta\phi^{*}_{q}-\frac{1}{2}\eta^{\rho\sigma}\partial_{\rho}\delta\phi^{*}_{p}\partial_{\sigma}\delta\phi^{*}_{q}\\
   &-\xi\left(\partial_{0}\partial_{0}-\frac{a'}{a}\partial_{0}-\eta^{\rho\sigma}\partial_{\rho}\partial_{\sigma}-3\left(\frac{a'}{a}\right)^{2}\right)\delta\phi^{*}_{p}\delta\phi^{*}_{q}\biggr] \\
   &\times e^{-i\left(\textbf{p}+\textbf{q}\right)\cdot \textbf{x}},\label{A0'}
\end{split}
\end{equation}
and
\begin{equation}
\begin{split}   &A_{i}\left(\textbf{x},\tau\right)=2\Psi\biggl[\partial_{i}\delta\phi^{*}_{p}\partial_{i}\delta\phi^{*}_{q}+\frac{1}{2}\eta^{\rho\sigma}\partial_{\rho}\delta\phi^{*}_{p}\partial_{\sigma}\delta\phi^{*}_{q}\\
   &-\xi\left(\partial_{i}\partial_{i}+3\frac{a'}{a}\partial_{0}+\eta^{\rho\sigma}\partial_{\rho}\partial_{\sigma}+2\frac{a''}{a}-\left(\frac{a'}{a}\right)^{2}\right)\delta\phi^{*}_{p}\delta\phi^{*}_{q}\biggr]\\
&\times e^{-i\left(\textbf{p}+\textbf{q}\right)\cdot \textbf{x}}.\label{Ai'}
\end{split}
\end{equation}
In analogy to Eq. \eqref{meanback}, during slow-roll we  approximate
\begin{align}
   \phi\rightarrow \frac{\int ^{\tau_{f}}_{\tau_{i}}\phi d\tau}{\tau_{f}-\tau_{i}}\equiv \braket{\phi},\quad
   \phi^\prime \rightarrow  \frac{\int ^{\tau_{f}}_{\tau_{i}}\phi'd\tau}{\tau_{f}-\tau_{i}} \equiv \braket{\phi^\prime},
\end{align}
that turn out to be suitable replacement in order to obtain analytical results for the total amplitude of particle production.

Such probability amplitude can be then computed following the same prescriptions for super-Hubble modes discussed in Sec. \ref{Sez3A}. As in small-field inflation, the total amplitude requires proper normalization with respect to the zero-order Lagrangian associated to inflaton fluctuations.

It is evident that in the case of large-field inflation, as expressed by the Lagrangian in Eq.~\eqref{LagrLargefield}, there is the absence of a constant term. Consequently, the ``cosmological constant" undergoes evolution during the inflationary period, aligning with our previous expectations. In other words, the cosmological constant is no longer constant.
However, this is plausible, at least for two main reasons, say
\begin{itemize}
    \item[-] the phase in which the cosmological constant is no longer a pure constant represents a \emph{metastable case}. It is associated with a phase transition and involves the presence of perturbations, that inevitably break the original translational invariance. Accordingly, Weinberg's no-go theorem \cite{CCP0} is not expected to hold there, and no fine-tuning is required for the cosmological constant.
    \item[-] at the end of inflation, the constant contribution is tuned by  the mechanism of phase transition and therefore is restored. At this point, inhomogeneities are typically damped out by backreaction  and/or decoherence mechanisms \cite{backr}. So, there is no more violation to the no-go theorem, after the transition.
\end{itemize}

The varying cosmological constant value can be denoted as $\Lambda^{4}\left(\phi\right)=\chi\phi^{4}/4$ and appears similar to some cases developed in the literature \cite{sola1,sola2,sola3}, albeit severely circumscribed within the phase transition only and not elsewhere. Consequently, the Hubble parameter during inflation is given by
\begin{equation}
    H^{2}_{I}=\frac{8\pi G}{3}\rho_{\phi},\label{HubbleNewInf}
\end{equation}
where the corresponding inflaton energy density is determined as
\begin{equation}
    \rho_{\phi}=\frac{1}{2}\dot{\phi}^{2}+\frac{\chi}{4}\phi^{4}+\frac{1}{2}\xi R\phi^{2}.
\end{equation}

In slow-roll regime, $\dot{\phi}^{2}\ll V^{\rm eff}$, the energy density thus becomes
\begin{equation}
    \rho_{\phi}=\frac{\chi}{4}\phi^{4}+\frac{1}{2}\xi R\phi^{2}=\frac{\chi}{4}\phi^{4}+\frac{1}{2}\xi \left(8\pi G T^{\mu}_{\mu}\right)\phi^{2},
\end{equation}
having $T^{\mu}_{\mu}=\xi R\phi^{2}+\chi \phi^{4}$ and
\begin{equation}
    R=\frac{8 \pi G}{1-8\pi G\xi\phi^{2}}\chi \phi^{4}.
\end{equation}

After the phase transition, the field screens also the gravitational constant $G$.

However, in the large-field case we have $\braket{\phi}^{2}\gg v^{2}$, so $ 8\pi G\xi v^{2}\ll 1$ for realistic values of the coupling constant $\xi$. In this way, Eq. \eqref{constra} can be easily satisfied and Einstein field equations are restored without requiring the fine-tuning $\xi \leq 10^{-5}$.

\begin{figure}[H]
        \includegraphics[width=8cm]{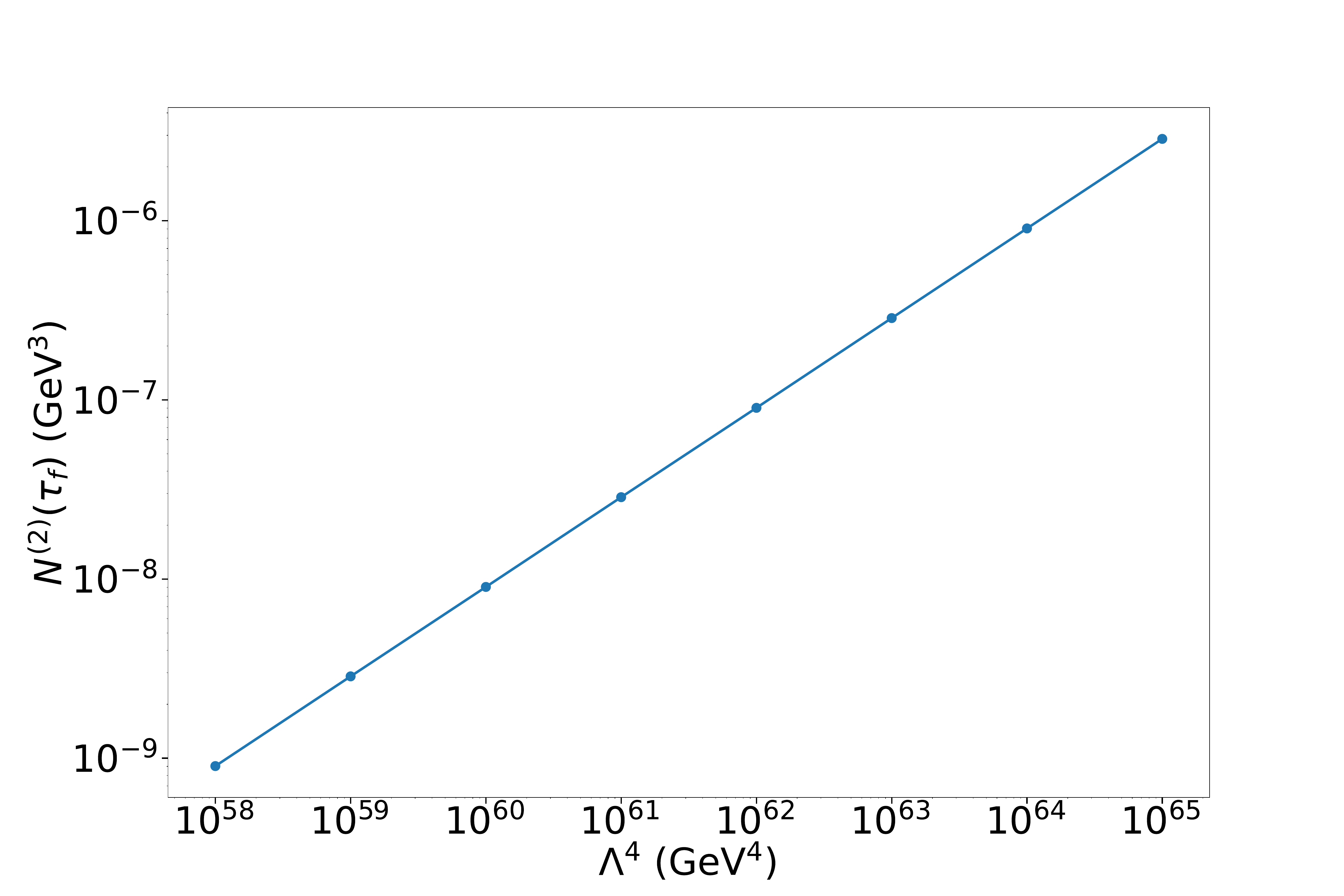}
    \caption{Number density $N^{(2)}$ in $\unit{\giga\electronvolt}^{3}$ as function of vacuum energy $\Lambda^{4}$ in loglog scale. The number density is computed assuming $N=60$ e-foldings, $\tau_{i}=-10^{3}$ \unit{\giga\electronvolt}$^{-1}$, $\tau'=-5\times 10^{-7}$ \unit{\giga\electronvolt}$^{-1}$, $\epsilon=10^{-3}$ and $\xi=10^{-3}$.}
    \label{VarLlarge1}
\end{figure}

Fig.~\ref{VarLlarge1} shows the number density of particles produced during inflation as function of vacuum energy, $\Lambda^{4}$, assuming $\tau^\prime=-5\times10^{-7}$ \unit{\giga\electronvolt}$^{-1}$.

\begin{figure}[H]
\begin{center}
\includegraphics[width=8cm]{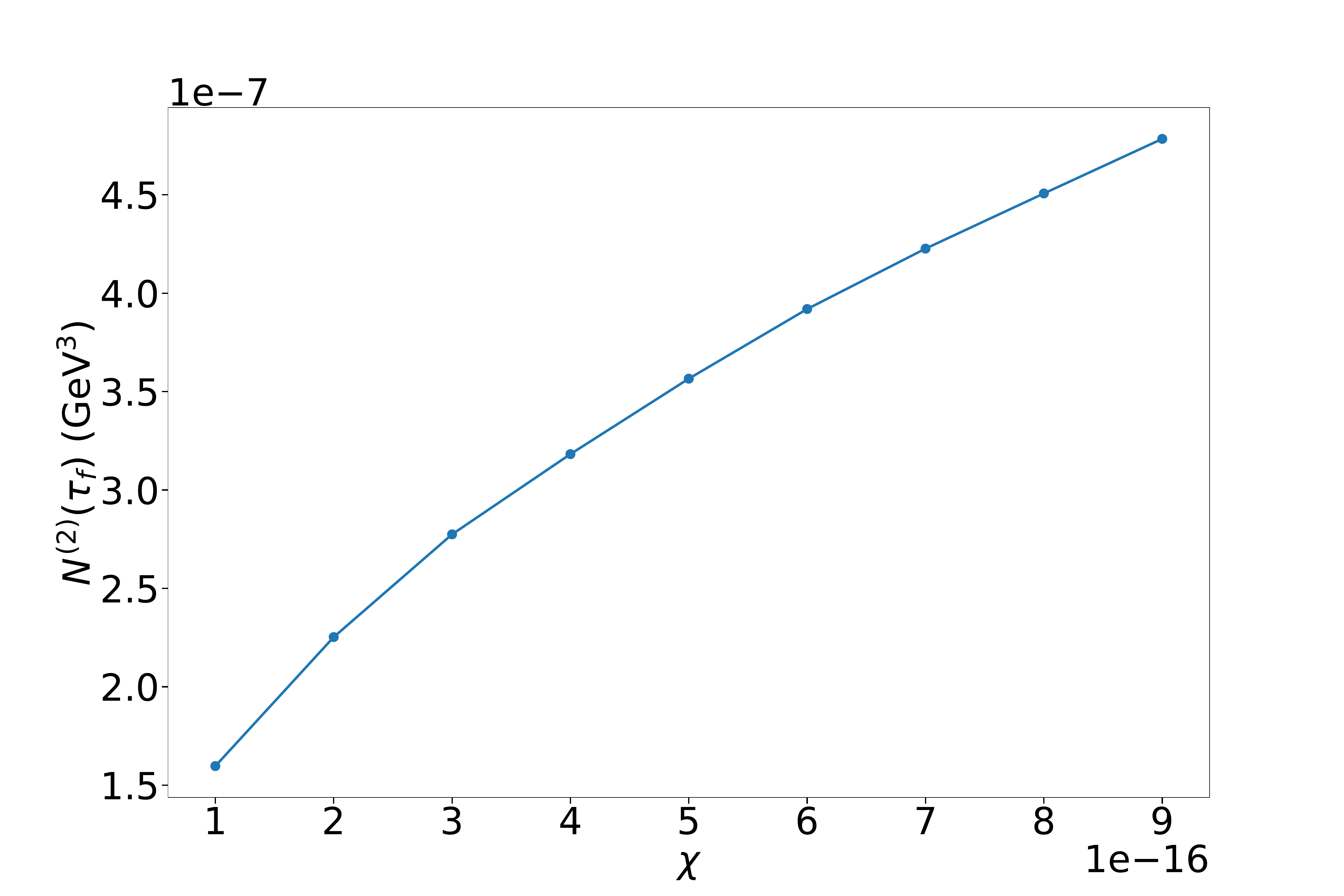}
\caption{Number density $N^{(2)}$ in $\unit{\giga\electronvolt}^{3}$ as function of the
self-coupling constant $\chi$, with $\tau^\prime=-5\times 10^{-7}$ \unit{\giga\electronvolt}$^{-1}$. In this figure, $\chi$ is chosen so to
respect the constraint $\sqrt{\chi/\xi^2} \sim 10^{-5}$, namely Eq. \eqref{constra}. The other parameters are the same as in Fig. \ref{VarLlarge1}.}\label{LargeChiVar2}
\end{center}
\end{figure}

In Fig.~\ref{LargeChiVar2}, the same number density is depicted as function of self-coupling constant $\chi$, selecting the range in which the constraint of Eq. \eqref{constra} for non-minimal coupling inflation is fulfilled, as previously discussed. The values adopted to display out plots are reported in Tabs.~\ref{Tab2}-\ref{Tab3}.

\begin{table}[H]
\begin{center}
\begin{tabular}{||c | c | c | c||}
\hline
$\Lambda^{4}$ (\unit{\giga\electronvolt})$^{4}$ & $\chi$ & $N^{(2)}\left(\tau_{f}\right)$
 (\unit{\giga\electronvolt})$^{3}$\\ [0.5ex]
 \hline\hline
 $\num{1.0}\times 10^{58}$ & $\num{3.2}\times 10^{-21}$  & $\num{9.0}\times 10^{-10}$  \\
 $\num{1.0}\times 10^{59}$&  $\num{3.2}\times 10^{-20}$ & $\num{2.9}\times 10^{-9}$ \\
$\num{1.0}\times 10^{60}$  &  $\num{3.2}\times 10^{-19}$  & $\num{9.0}\times 10^{-9}$\\
$\num{1.0}\times 10^{61}$ & $\num{3.2}\times 10^{-18}$  & $\num{2.9}\times 10^{-8}$\\
 $\num{1.0}\times 10^{62}$ & $\num{3.2}\times 10^{-17}$  & $\num{9.0}\times 10^{-8}$ \\
 $\num{1.0}\times 10^{63}$ & $\num{3.2}\times 10^{-16}$  & $\num{2.9}\times 10^{-7}$ \\
 $\num{1.0}\times 10^{64}$   & $\num{3.2}\times 10^{-15}$  & $\num{9.0}\times 10^{-7}$ \\
 $\num{1.0}\times 10^{65}$ & $\num{3.2}\times 10^{-14}$ & $\num{2.9}\times 10^{-6}$ \\ \hline
\end{tabular}
\end{center}

\caption{Number density of particles produced at different cut-off times, for given values of vacuum energy and corresponding self-coupling constant.}
\label{Tab2}
\end{table}

\begin{table}[H]
\begin{center}
\begin{tabular}{||c | c | c||}
\hline
$\Lambda^{4}$ (\unit{\giga\electronvolt})$^{4}$& $\chi$ & $N^{(2)}\left(\tau_{f}\right)$
 (\unit{\giga\electronvolt})$^{3}$\\ [0.5ex]
 \hline\hline
 $\num{3.1}\times 10^{62}$&$\num{1.0}\times 10^{-16}$  & $\num{1.6}\times 10^{-7}$ \\
$\num{6.2}\times 10^{62}$&$\num{2.0}\times 10^{-16}$ & $\num{2.3}\times 10^{-7}$ \\
$\num{9.3}\times 10^{62}$&$\num{3.0}\times 10^{-16}$ & $\num{2.8}\times 10^{-7}$\\
$\num{1.2}\times 10^{63}$&$\num{4.0}\times 10^{-16}$ & $\num{3.2}\times 10^{-7}$ \\
$\num{1.6}\times 10^{63}$&$\num{5.0}\times 10^{-16}$ & $\num{3.6}\times 10^{-7}$ \\
$\num{1.9}\times 10^{63}$&$\num{6.0}\times 10^{-16}$ & $\num{3.9}\times 10^{-7}$ \\
$\num{2.2}\times 10^{63}$&$\num{7.0}\times 10^{-16}$ & $\num{4.2}\times 10^{-7}$ \\
$\num{2.5}\times 10^{63}$&$\num{8.0}\times 10^{-16}$ & $\num{4.5}\times 10^{-7}$ \\
 $\num{2.8}\times 10^{63}$&$\num{9.0}\times 10^{-16}$  & $\num{4.8}\times 10^{-7}$  \\
\hline
\end{tabular}
\end{center}
\caption{Number density of particles produced assuming $\sqrt{\chi/\xi^2} \sim 10^{-5}$ and $\tau^\prime=-5\times 10^{-7}$ \unit{\giga\electronvolt}$^{-1}$.
The vacuum energy contribution $\Lambda^4$ is chosen so to keep fixed the average value of inflaton field $\langle \phi \rangle$ in slow-roll at any given $\chi$, with initial condition $\phi\left(\tau_i\right)=5\  M_{p}$.}
\label{Tab3}
\end{table}

 We notice that, as the self-coupling constant
$\chi$ increases, the magnitude of vacuum energy $\Lambda^{4}=\chi\phi^{4}/4$ is also larger. Since the inflaton field remains approximately constant during slow-roll, vacuum energy can be amplified solely by increasing the self-coupling constant, leading to a corresponding enhancement of particle production.


\section{Contribution to dark matter magnitude}\label{sezione5}

Geometric particles produced during inflation could admit an interpretation in terms of dark matter \cite{DM1, DM2, geocorr}. Dark matter particles of gravitational origin have been recently proposed in several scenarios, usually assuming scalar or vector fields, but also considering higher-spin candidates \cite{dmg1,dmg2,dmg3,dmg4, dmg5, dmg6}.
We here suggest that dark matter may arise from the perturbative approach previously discussed, where the Yukawa-like interaction term $\frac{\xi}{2}R\phi^{2}$ ‘dresses' the inflaton field $\phi$ by the interaction. Examples of such a model are also possible in condensed matter, see e.g. \cite{condensed}, and in gravitational contexts \cite{gwparticles,altroparticles}.

Under this interpretation, the corresponding particles look like geometric quasi-particles, representing excitations of the inflaton field and the scalar curvature. As these particles appear as a consequence of the non-minimal coupling, we could expect their spins to span within the interval $[0;2]$, where $0$ corresponds to the scalar field spin, whereas $2$ to gravitons. The fundamental nature of such contribution might be investigated analyzing how much and whether these objects appear stable, working out their lifetimes. Specifically, in view of our results, we can easily state here that geometric particle production typically leads to \emph{particle pairs with different momenta}, representing the main difference with respect to purely gravitational production. In this latter case, the sole expansion of the universe is responsible for the process, implying the creation of particle-antiparticle pairs with same momenta.

Nonperturbative particle production from vacuum can be quantified by means of the Bogoliubov coefficients relating \emph{in} and \emph{out} vacuum states of the system. In particular, gravitational production during inflation has been proposed in recent years for both fermionic and bosonic dark matter candidates \cite{dmg1,dmg2,dmg3,dmg4,dmg5,dmg6}. Despite such nonperturbative production from vacuum should result in larger number densities with respect to perturbative geometric processes, the creation of pairs with equal and opposite momenta may lead in principle to annihilation. Moreover, in case of conformally coupled field candidates, nonperturbative gravitational production is typically negligible \cite{boya}, thus necessarily requiring different approaches. Considering this, for the moment we disregard the mentioned pairs, assuming that they undergo annihilation before making a significant contribution to the abundance of dark matter. A comparison between gravitational production from vacuum and geometric mechanisms of particle production is left for future investigations.  Consequently, we now proceed to quantify the mass of our proposed candidate for geometric dark matter, simplifying for the moment the analysis by assuming negligible Bogoliubov coefficients related to nonperturbative processes\footnote{As discussed in Appendix \ref{AppC}, nonzero Bogoliubov coefficients may also enhance the geometric mechanism of production, thus resulting in a larger number of geometric particles produced during slow-roll. We plan to come back to this point in future works, so to include such contribution in our treatment.}.

After the reheating epoch, the Hubble parameter in the radiation-dominated era is given by
\begin{equation}
    H(z)^{2}\simeq H_{0}^{2}\Omega_{r,0}\left(1+z\right)^{4},\label{Hzred}
\end{equation}
where $\Omega_{r,0}$ is the current radiation energy density. During the radiation phase, the Hubble parameter is related to the temperature as $H^{2}\sim G T^{4}$, implying that the redshift $z$ at the beginning of radiation-dominated era can be determined as
\begin{equation}
    z\left(T_{r}\right)\simeq \left(\frac{G T_{r}^{4}}{H_{0}^{2}\Omega_{r,0}}\right)^{\frac{1}{4}},\label{redshift}
\end{equation}
where $T_{r}$ denotes the temperature at the end of reheating.

Assuming that all dark matter was generated during inflation via the geometric mechanism previously described, we can estimate its mass $m^*$ from the corresponding number density, namely
\begin{equation}
\rho_{DM} = m^* N^{(2)}.
\end{equation}
Thus, we obtain
\begin{equation} \label{masdm}
    m^{*}(T_{r})=\frac{\rho_{DM}}{N^{(2)}}\simeq \left(\frac{G T_{r}^{4}}{H_{0}^{2}\Omega_{r,0}}\right)^{\frac{3}{4}}\frac{\rho_{DM,0}}{N^{(2)}},
\end{equation}
since
$\rho_{DM}\left(\tau_{r}\right)=\rho_{DM,0}\left(1+z\right)^{3}$ holds for dark matter. The number density $N^{(2)}$ is given by Eq.~\eqref{GNP}, where the probability amplitude for particle production has been previously derived in both limits of small and large-field inflation. In principle, the number density might be computed at the end of preheating, say at $\tau_r$. However, we can assume that the scale factor does not vary significantly from the end of inflation to the first stage of radiation-dominated phase, thus setting $a(\tau_{f})=a(\tau_{r})$.

 Moreover, from Eq. \eqref{masdm} we notice that the mass of our dark matter candidate turns out to be a model-dependent quantity, related to the final temperature of the reheating process\footnote{In doing so, we are neglecting the effects of reheating on particle production. This assumption seems realistic in case of super-Hubble modes, since wavelengths larger than the Hubble horizon are causally disconnected from the microphysical processes of thermalization taking place after inflation. However, if the reheating temperature is sufficiently low (e.g. a few MeV \cite{lowreh}), superhorizon modes may reenter before the beginning of the radiation phase, thus modifying the overall picture \cite{boya}.}.

For the sake of clearness, constructing new particles implies the existence of a non-zero equation of state. The latter should resemble that of dark matter, in order to guarantee that our particles are associated with such a species. Imposing the condition that at reheating the main contribution is the one discussed above, we can bound the equation of state itself to show a matter behavior at the end of inflation. Throughout the inflationary epoch, however, the equation of state is no longer zero, i.e., it is not dust, but rather it is provided by computing the standard energy-momentum tensor, $T_{\mu\nu}\equiv\frac{\delta S}{\delta g_{\mu\nu}}$. Hence, during inflation we expect matter with pressure to mime dark matter that, once inflation ends, turns out to diluted, tending to dust, namely to zero. These features, as well as stability of dark matter particles produced from the non-minimal coupling, will be subject of future investigations.


\subsection{Small-field domain}

\begin{figure}[H]
    \begin{subfigure}{0.5\textwidth}
        \includegraphics[width=8cm]{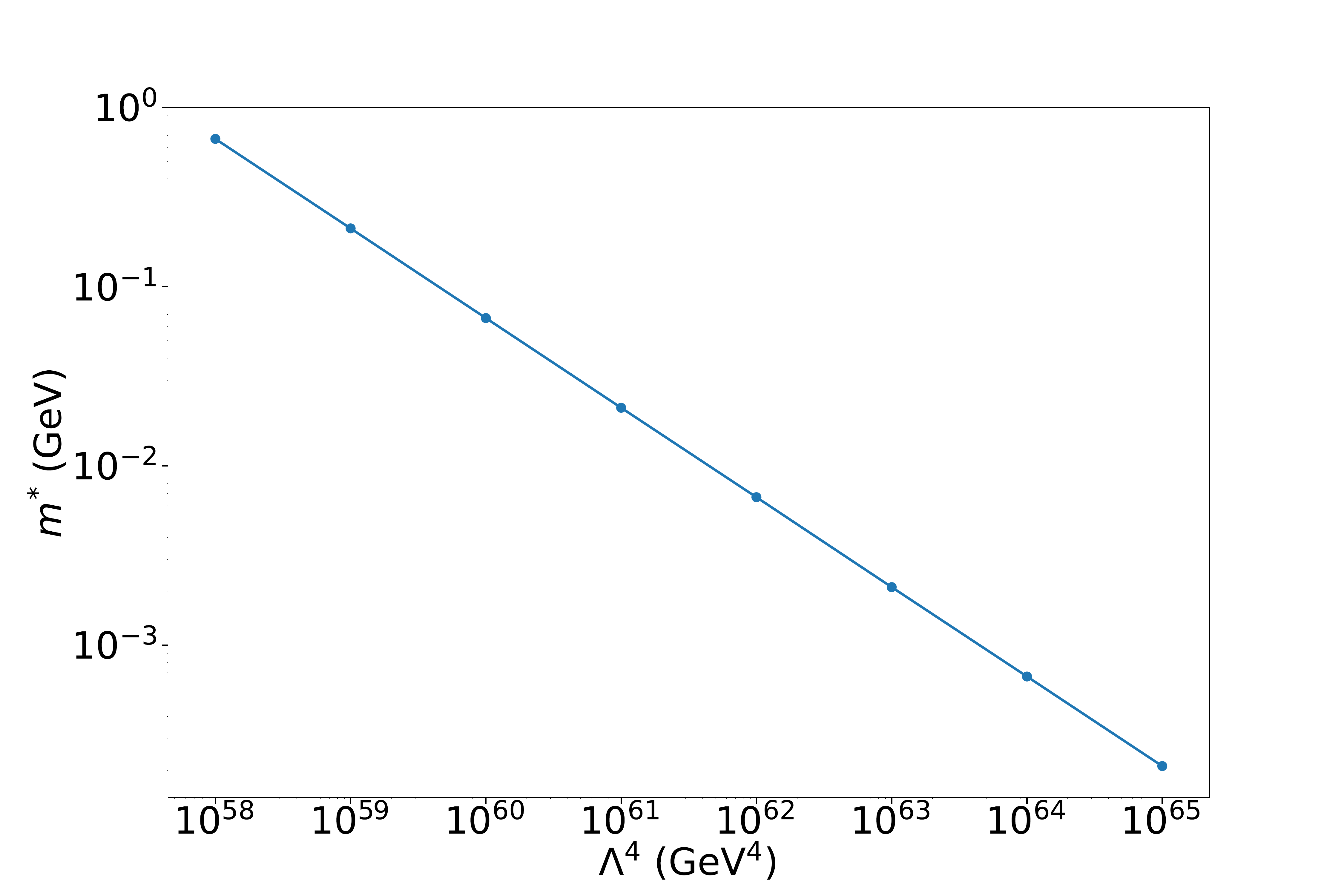}
    \end{subfigure}
    \caption{Mass $m^{*}$ of the dark matter candidate
as function of vacuum energy $\Lambda^{4}$ in loglog scale with initial condition $\phi(\tau_i)=v/10^{20}$. The number density is computed assuming $a(\tau_{f})=a(\tau_{r})$ and the super-Hubble modes are chosen under the prescription $\tau_{i}=-10^{3}$ \unit{\giga\electronvolt}$^{-1}$ and $\tau^\prime=-5\times10^{-2}$  \unit{\giga\electronvolt}$^{-1}$. The other parameters are: $\epsilon=10^{-3}$, $\xi=10^{-5}$ and $T_r=1$ GeV.}
\label{mLVarTr1}
    \begin{subfigure}{0.5\textwidth}
        \includegraphics[width=8cm]{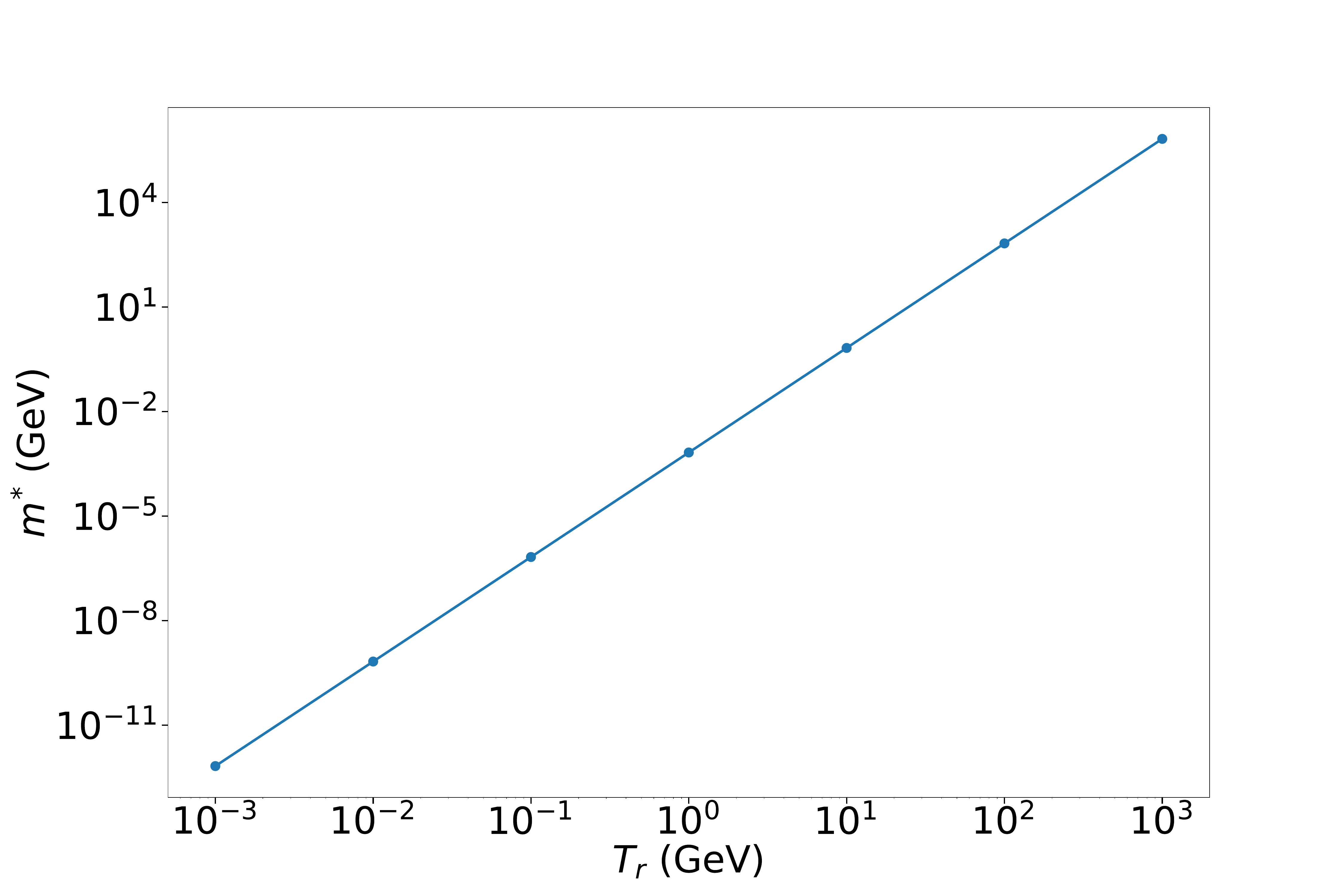}
    \end{subfigure}
    \caption{Mass $m^{*}$
as function of temperature $T_{r}$ in loglog scale with vacuum energy $\Lambda^{4}=10^{64}$ $\unit{\giga\electronvolt}^{4}$ and self-coupling constant $\chi=10^{-14}$. The other parameters are the same as in Fig. \ref{mLVarTr1}.}
\label{massL1064Trvario}
\end{figure}

In Fig.~\ref{mLVarTr1} we show the mass of dark matter candidate at fixed temperature, $T_{r}=1$ \unit{\giga\electronvolt}, as function of vacuum energy $\Lambda^{4}$. Instead, in Fig.~\ref{massL1064Trvario} the mass of dark matter candidate is plotted at fixed vacuum energy, $\Lambda^{4}=10^{64}$ $\unit{\giga\electronvolt}^{4}$, as function of temperature $T_{r}$. Finally, in Tabs.~\ref{Tab4} and~\ref{Tab5} dark matter mass values are synthesized as function of vacuum energy $\Lambda^{4}$ and temperature $T_{r}$, respectively.

\begin{table}[H]
\begin{center}
\begin{tabular}{||c | c | c | c||}
\hline
$\Lambda^{4}$ (\unit{\giga\electronvolt})$^{4}$  & $\chi$ & $N^{(2)}\left(\tau_{r}\right)$
 (\unit{\giga\electronvolt})$^{3}$ & $m^{*}$ (\unit{\giga\electronvolt})\\ [0.5ex]
 \hline\hline
$\num{1.0}\times 10^{58}$ & $1.0\times 10^{-20}$ & $\num{2.0}\times 10^{-10}$ & $6.7 \times 10^{-1}$  \\
$\num{1.0}\times 10^{59}$& $1.0\times 10^{-19}$ &  $\num{6.3}\times 10^{-10}$ & $2.1 \times 10^{-1}$\\
$\num{1.0}\times 10^{60}$ & $1.0\times 10^{-18}$ & $\num{2.0}\times 10^{-9}$ & $6.7 \times 10^{-2}$ \\
$\num{1.0}\times 10^{61}$ & $1.0\times 10^{-17}$ & $\num{6.3}\times 10^{-9}$ & $2.1 \times 10^{-2}$ \\
$\num{1.0}\times 10^{62}$ & $1.0\times 10^{-16}$ & $\num{2.0}\times 10^{-8}$ & $6.7 \times 10^{-3}$\\
$\num{1.0}\times 10^{63}$ & $1.0\times 10^{-15}$ & $\num{6.3}\times 10^{-8}$ & $2.1 \times 10^{-3}$ \\
$\num{1.0}\times 10^{64}$   & $1.0\times 10^{-14}$ & $\num{2.0}\times 10^{-7}$ & $6.7 \times 10^{-4}$ \\
$\num{1.0}\times 10^{65}$ & $1.0\times 10^{-13}$ & $\num{6.3}\times 10^{-7}$ & $2.1 \times 10^{-4}$\\
\hline
\end{tabular}
\end{center}
\caption{Mass of the dark matter candidate for different values of vacuum energy $\Lambda^4$ and self-coupling constant $\chi$, with $T_{r}=1$ \unit{\giga\electronvolt} and $\tau^\prime=-5\times 10^{-2}$ \unit{\giga\electronvolt}$^{-1}$.}
\label{Tab4}
\end{table}

\begin{table}[H]
\begin{center}
\begin{tabular}{||c | c||}
\hline
 $T_{r}$ (\unit{\giga\electronvolt}) & $m^{*}$
 (\unit{\giga\electronvolt}) \\ [0.5ex]
 \hline\hline
\num{1.0}$\times 10^{-3}$ &\num{6.7}$\times 10^{-13}$\\
\num{1.0}$\times 10^{-2}$  & \num{6.7}$\times 10^{-10}$\\
\num{1.0}$\times 10^{-1}$  & \num{6.7}$\times 10^{-7}$\\
\num{1.0}  &  \num{6.7}$\times 10^{-4}$\\
\num{1.0}$\times 10^{1}$   & \num{6.7}$\times 10^{-1}$\\
\num{1.0}$\times 10^{2}$ &  \num{6.7}$\times 10^{2}$\\
\num{1.0}$\times 10^{3}$  & \num{6.7}$\times 10^{5}$\\ \hline
\end{tabular}
\end{center}

\caption{Mass of the dark matter candidate for different values of the temperature $T_{r}$. We assume $\tau^\prime=-5\times 10^{-2}$ \unit{\giga\electronvolt}$^{-1}$, $\Lambda^{4}=10^{64}$ $\unit{\giga\electronvolt}^{4}$, $\chi=10^{-14}$ and $N^{(2)}(\tau_{r})=9.1\times 10^{-7}$ $\unit{\giga\electronvolt}^{3}$.}
\label{Tab5}
\end{table}

\subsection{Large-field domain}

\begin{figure}[H]
        \includegraphics[width=8cm]{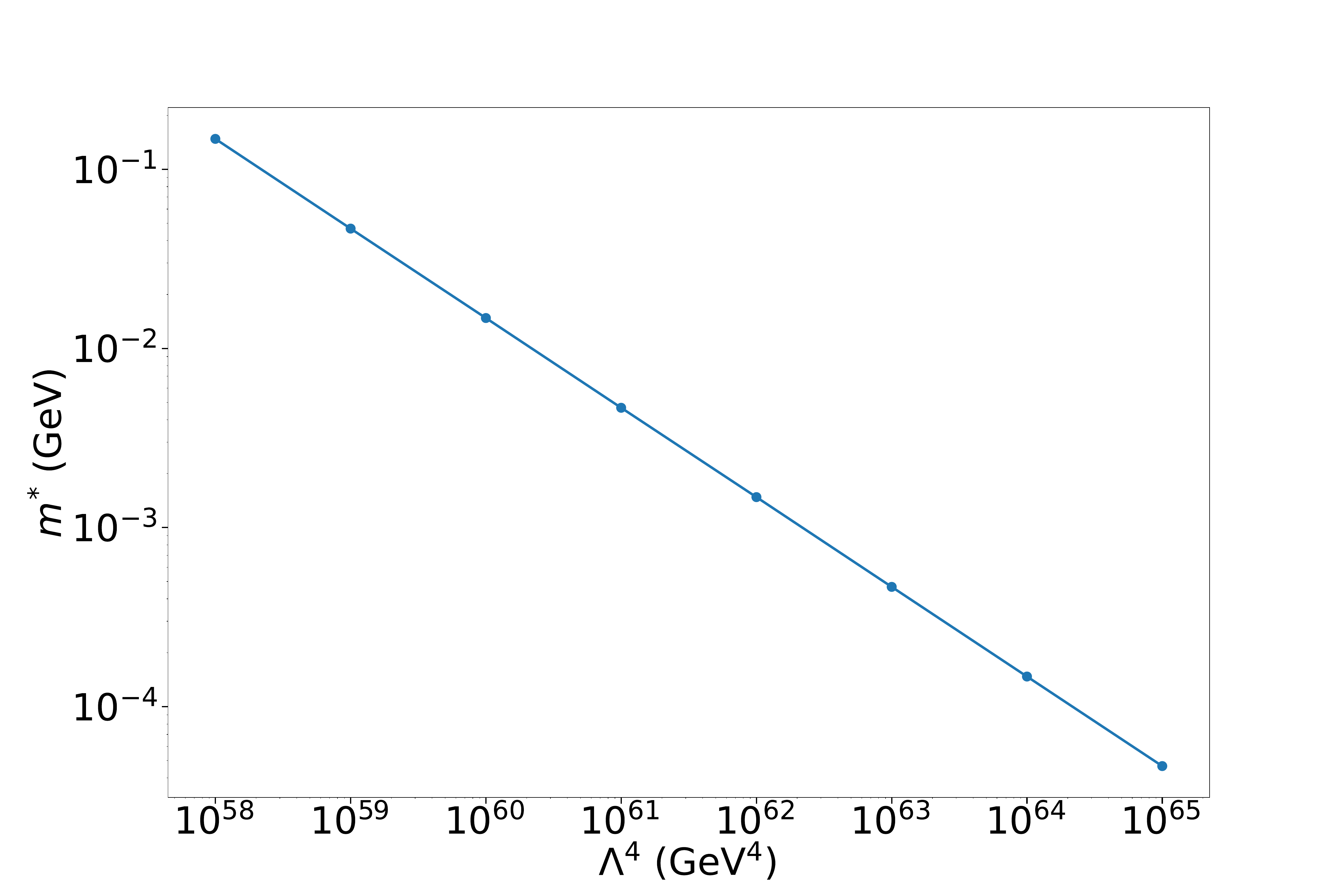}
\caption{Mass $m^{*}$ of the dark matter candidate
as function of vacuum energy $\Lambda^{4}$ in loglog scale with initial condition $\phi\left(\tau_{i}\right)=5 M_{P}$. The number density is computed assuming $a(\tau_{f})=a(\tau_{r})$ and the super-Hubble modes are chosen under the prescription $\tau_{i}=-10^{3}$ \unit{\giga\electronvolt}$^{-1}$ and $\tau^\prime=-5\times10^{-7}$  \unit{\giga\electronvolt}$^{-1}$. The other parameters are: $\epsilon=10^{-3}$, $\xi=10^{-3}$ and $T_{r}=1$ \unit{\giga\electronvolt}.}
\label{LargeDMTr1}
\end{figure}

\begin{figure}[H]
        \includegraphics[width=8cm]{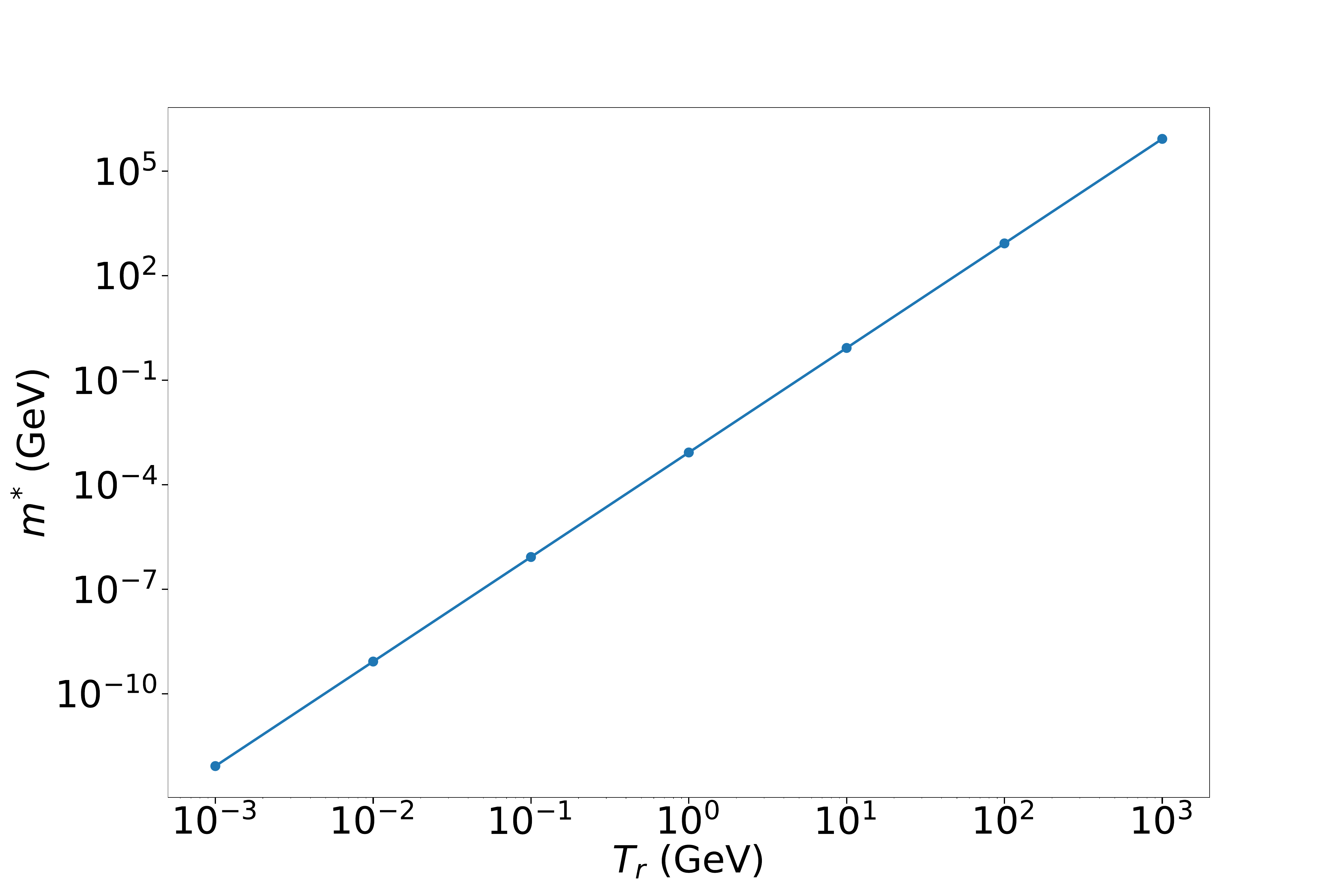}
    \caption{ Mass $m^{*}$
as function of temperature $T_{r}$ in loglog scale with vacuum energy $\Lambda^{4}=3.1\times10^{62}$ $\unit{\giga\electronvolt}^{4}$ and self-coupling constant $\chi=10^{-16}$. The other parameters are the same as in Fig. \ref{LargeDMTr1}.}
\label{LargeTrvar}
\end{figure}

The same calculations have been performed for the large-field scenario. In particular, Fig.~\ref{LargeDMTr1} shows the mass of dark matter candidate at $T_{r}=1$ \unit{\giga\electronvolt}, as function of vacuum energy $\Lambda^{4}$. Fig.~\ref{LargeTrvar} exhibits how the particle mass varies with respect to the temperature at fixed vacuum energy. Tabs.~\ref{Tab6} and~\ref{Tab7} provide a summary of the mass of dark matter candidate, as a function of vacuum energy $\Lambda^{4}$ and temperature $T_{r}$, respectively.


\begin{table}[H]
\begin{center}
\begin{tabular}{||c | c | c | c||}
\hline
$\Lambda^{4}$ (\unit{\giga\electronvolt})$^{4}$ &  $\chi$ & $N^{(2)}\left(\tau_{r}\right)$
 (\unit{\giga\electronvolt})$^{3}$ & $m^{*}$
 (\unit{\giga\electronvolt}) \\ [0.5ex]
 \hline\hline
$\num{1.0}\times 10^{58}$ & $\num{3.2}\times 10^{-21}$  & $\num{9.0}\times 10^{-10}$ & $1.0\times 10^{-1}$  \\
$\num{1.0}\times 10^{59}$&  $\num{3.2}\times 10^{-20}$ & $\num{2.9}\times 10^{-9}$ & $5.0\times 10^{-2}$  \\
$\num{1.0}\times 10^{60}$  &  $\num{3.2}\times 10^{-19}$  & $\num{9.0}\times 10^{-9}$ & $1.0\times 10^{-2}$\\
$\num{1.0}\times 10^{61}$ & $\num{3.2}\times 10^{-18}$  & $\num{2.9}\times 10^{-8}$ & $5.0\times 10^{-3}$\\
$\num{1.0}\times 10^{62}$ & $\num{3.2}\times 10^{-17}$  & $\num{9.0}\times 10^{-8}$ & $1.0\times 10^{-3}$ \\
$\num{1.0}\times 10^{63}$ & $\num{3.2}\times 10^{-16}$  & $\num{2.9}\times 10^{-7}$ & $5.0\times 10^{-4}$ \\
$\num{1.0}\times 10^{64}$   & $\num{3.2}\times 10^{-15}$  & $\num{9.0}\times 10^{-7}$ & $1.0\times 10^{-4}$\\
$\num{1.0}\times 10^{65}$ & $\num{3.2}\times 10^{-14}$ & $\num{2.9}\times 10^{-6}$ & $5.0\times 10^{-5}$\\ \hline
\end{tabular}
\end{center}
\caption{Mass of dark matter candidate as function of the temperature $T_{r}=1$ \unit{\giga\electronvolt}, with $\tau^\prime=-5\times 10^{-7}$ $\unit{\giga\electronvolt}^{-1}$.}
\label{Tab6}
\end{table}

\begin{table}[H]
\begin{center}
\begin{tabular}{||c | c||}
\hline
 $T_{r}$ (\unit{\giga\electronvolt}) & $m^{*}$
 (\unit{\giga\electronvolt}) \\ [0.5ex]
 \hline\hline
\num{1.0}$\times 10^{-3}$ &\num{8.4}$\times 10^{-13}$\\
\num{1.0}$\times 10^{-2}$  & \num{8.4}$\times 10^{-10}$\\
\num{1.0}$\times 10^{-1}$  & \num{8.4}$\times 10^{-7}$\\
\num{1.0}  &  \num{8.4}$\times 10^{-4}$\\
\num{1.0}$\times 10^{1}$ &  \num{8.4}$\times 10^{-1}$\\
\num{1.0}$\times 10^{2}$ &  \num{8.4}$\times 10^{2}$\\
\num{1.0}$\times 10^{3}$  & \num{8.4}$\times 10^{5}$\\ \hline
\end{tabular}
\end{center}
\caption{Mass of the dark matter candidate varying the temperature $T_{r}$, with $\tau^\prime=-5\times 10^{-7}$ $\unit{\giga\electronvolt}^{-1}$, $\Lambda^{4}=3.1\times10^{62}$ $\unit{\giga\electronvolt}^{4}$, $N^{(2)}(\tau_{r})=1.6\times 10^{-7}$ $\unit{\giga\electronvolt}^{3}$ and $\chi=10^{-16}$.}
\label{Tab7}
\end{table}

\section{Comparison between small and large field particle production}\label{sezione6}

Analyzing the symmetry breaking potential of Eq. \eqref{effpot} in the limit of small and large-field inflation, we have shown that both approaches allow to produce particles arising from inflaton fluctuations. We also notice that the offset term $V_0$ does not affect the total amount of particles produced. However, there are important differences in the choice of the parameters:
\begin{itemize}
    \item[-] in the small-field scenario, vacuum energy is described by the constant term $\Lambda^{4}=\chi v^{4}/4$, so it essentially depends on the field value at the minimum of the potential. If $v$ is chosen to be of the order of Planck mass, then in order to satisfy $8\pi G v^2 \ll 1$ we need $\xi \leq 10^{-5}$. However, this condition violates the requirement of Eq. \eqref{constra};
    \item[-] in the large-field scenario, vacuum energy is denoted by $\Lambda^{4}=\chi \braket{\phi}^{4}/4$, so it is independent from the value of the minimum $v$. Instead, it is related to the value of the inflaton field during slow-roll. In this case $\braket{\phi}\simeq 10^{19}$ \unit{\giga\electronvolt} for all values of $\chi$ used in the work. Since in large-field inflation we require $\phi \gg v$ during slow-roll, we can safely satisfy the condition of negligible screening of Newton's constant at the end of inflation, namely $8 \pi G \xi v^2 \ll 1$, respecting at the same time the constraint given by Eq. \eqref{constra}.
\end{itemize}
We then notice that large-field inflation is favored over the small-field approach when dealing with a symmetry breaking potential, since it allows to satisfy the constraint of Eq. \eqref{constra} without significantly affect Newton's gravitational constant at the end of inflation. At the same time, even if the small-field scenario can still produce a relevant number of geometric particles during slow-roll, it violates the condition \eqref{constra} if we require a negligible screening of Newton's gravitational constant due to field-curvature coupling.

For what concerns the mass of the dark matter candidate, we observe that in both scenarios a larger vacuum energy term during inflation implies smaller values for the mass of the geometric quasi-particles produced. Similarly, a larger self-coupling constant increases the number density of dark matter particles produced, thus resulting in a smaller mass. As shown in Figs. \ref{mLVarTr1} and  \ref{LargeDMTr1}, typical mass values span from a few eV up to the GeV scale for a fixed temperature of $T_r=1$ GeV, in both limits of small and large fields. Instead, larger masses are obtained in case of larger reheating temperature, see Figs. \ref{massL1064Trvario} and \ref{LargeTrvar} for the small and large-field case, respectively.

Additionally, as previously stated, we remark that our estimates also depend on the cut-off time $\tau^\prime$ through which we study the evolution of super-Hubble fluctuations during slow-roll. Further details on this point are given in Appendix \ref{AppD}.

\subsection{Effective value of the cosmological constant after inflation}

We underlined above that the presence of a nonzero offset $V_0$ does not increase or decrease the total number of geometric particles produced. However, it contributes to the effective value of the cosmological constant at the end of inflation, as shown by Eq. \eqref{Latot}.

In our model, the choice $V_0=0$ implies that the potential term does not affect the cosmological constant \emph{after the phase transition}, except for curvature effects. This would result in a bare cosmological constant contribution at the end of inflation. A negative offset, which we considered in our large-field approach, may still represent a valid alternative, \emph{provided this contribution is canceled by some other mechanism}. In particular, in our treatment we do not consider the presence of a kinetic term associated to the scalar field \emph{before and after the phase transition}, focusing instead on the sole slow-roll phase, where the aforementioned term is clearly negligible.

However, a nonzero kinetic term can play a key role in canceling vacuum energy after phase transition, as proposed in some recent works. This reinforces the idea that a large-field model of inflation with nonzero offset may correctly reproduce the amount of vacuum energy still present after the phase transition. More specifically, in Refs. \cite{Dustwithpress0, Dustwithpress1}, it has been shown that if local shift symmetry holds, a scalar field describing \emph{a single fluid of matter with pressure} may cancel vacuum energy density through the kinetic contribution, that turns out to be constant \emph{before and after the phase transition itself}. In this case, the presence of a negative offset allows to avoid the coincidence problem and the corresponding matter fluid exhibits nonzero pressure, exhibiting as \emph{an emergent cosmological constant} that becomes dominant after the phase transition.

Analogously, in Ref. \cite{luoqq}, a constant kinetic term is associated with the evolution of a quasi-quintessence field before and after a small-field inflationary phase. There, the cancellation mechanism results into the difference between the pressure of such dark matter field (that coincides with the kinetic term before transition) and the potential offset. More recently, small and large-field inflationary scenarios have been compared in the framework of geometric particle production across the Hubble horizon and corresponding entanglement generation \cite{belsup}. In particular, the authors show that the large-field symmetry breaking potential of Eq. \eqref{potlarg} is able to produce a significant amount of particles also in case of sub-Hubble modes, and geometric production may lead to entanglement entropy among sub- and super-Hubble modes. This is not the case of small-field hilltop scenarios, where the number density of particles produced across the horizon and the corresponding entanglement are typically negligible. Entanglement generation in particle creation processes may help to understand the quantum properties of gravitationally produced particles: this is true in particular for dark matter, which is expected to weakly interact with the Standard Model fields and thus should be less involved in decoherence processes. Another issue of small-field models is that they typically require a significant fine-tuning of the initial conditions, as recently shown \cite{incond0,incond1}. For example, if the kinetic energy is too large or the initial inflaton value is slightly displaced from the maximum, then inflation may last much less than the required number of e-foldings \cite{incond2}.

Coming back to our scenario, the inclusion of an additional kinetic term seems natural in the context of reheating, when the inflaton field is expected to oscillate around its minimum.  We thus expect a nonzero kinetic term \emph{after the phase transition}, which may then contribute to the total energy density of the scalar field and can be involved in the here-described cancellation mechanism. This aspect warrants further investigation, and it will represent a key point in characterizing the post-inflationary dynamics within the framework that we have developed in our manuscript.

\section{Final remarks and perspectives}\label{sezione7}

We here examined inflation arising from a symmetry breaking potential coupled with curvature, considering both small and large inflaton fields. In so doing, we studied the evolution of inflaton fluctuations during slow-roll and showed that the corresponding potential reproduces inflation adopting a quasi-de Sitter phase. Specifically, we proposed to interpret the existence of inflation induced by a phase transition driven by vacuum energy aiming to address the cosmological constant problem.

Our proposal suggested that vacuum energy is effectively diminished as particles are generated during the phase transition. Rephrasing, this scenario highlights that the cosmological constant problem can be mitigated by converting the degrees of freedom associated with vacuum energy into the aforementioned matter particles.

Hence, we  quantified the amount of particles produced by metric perturbations in inflation, tracing them back to the inflaton fluctuations and showing how field-curvature coupling can enhance our mechanism of particle production. We interpreted the aforementioned particles as dark matter particles, and their abundance was calculated for both small and large inflaton fields. While both approaches yielded particle abundances that are consistent with observational constraints, it is only in the case of large fields, with the field minimum lying on Planck scales, that the current value of the gravitational constant can be accurately reproduced.

Furthermore, we emphasized the key distinctions between our geometric mechanism of particle production and the extensively-studied \emph{gravitational particle production}, which has been proposed as a mechanism for generating dark matter during inflation. In this regard, we proposed that the dark matter component may exist in the form of quasi-particles, arising from the coupling between particles and geometry, \emph{dressed} by the interaction between the inflaton and scalar curvature.

Accordingly, we computed mass and temperature at which particles arose in order to obtain the expected dark matter abundance. We also discussed the presence of an offset in the potential, showing that in our model the choice of a zero potential offset allows to obtain a bare cosmological constant at the end of inflation. However, a negative offset may represent a better ansatz if a further contribution determined from kinetic or potential energy of the scalar field is taken into account before and after phase transition. Moreover, we emphasized that vacuum energy ceases to remain constant during the phase transition, resulting in a varying effective cosmological constant. We thus discussed the physical interpretations of our findings, including its relation to the no-go theorem, as well as potential resolutions to overcome this issue. We then concluded that to fully-erase the cosmological constant, getting rid of the corresponding cosmological constant problem, additional degrees of freedom coming from kinetic and/or potential energy might be taken into account. Consequently, large-field inflation with non-minimal coupling appeared favored, being more compatible with theoretical expectations and, furthermore, guaranteeing to the Newton's constant to be compatible with observations after inflation.

In view of the above results, future developments will therefore focus on refining our treatment with the mechanism of vacuum energy cancellation presented in Refs. \cite{Dustwithpress0, Dustwithpress1}, where kinetic energy plays the role of reproducing the correct bare cosmological constant today.

For this reason, we plan to extend our study including post-inflationary dynamics, in order to provide a less approximate scenario after the phase transition. Finally, we will further investigate the fundamental properties of geometric quasi-particles, to show whether they can correctly reproduce dark matter abundance as here conjectured.

\vphantom{aaa}

\section*{Acknowledgements}

AB thanks the National Institute for Nuclear Physics for financial support. OL and YC acknowledge Marco Muccino for fruitful discussions toward the topic developed in this paper and, in particular, on reconciling the main subject of this work with the model previously presented. OL is also grateful to Carlo Cafaro, Stefano Mancini and Hernando Quevedo for interesting suggestions and comments.

\clearpage
\appendix

\section{Inflaton fluctuations in conformal time}\label{AppA}

We here show how to derive the dynamics of inflaton fluctuations in conformal time, namely assuming
\begin{equation}
g_{\mu\nu}=a^{2}(\tau)\eta_{\mu\nu}.
\end{equation}
In conformal time, field derivatives read
\begin{align}
&\dot{\phi}(t)=\frac{\phi'\left(\tau\right)}
{a\left(\tau\right)}\\
&\ddot{\phi}(t)=\frac{\phi''\left(\tau\right)}{a^{2}\left(\tau\right)}-\mathcal{H}\frac{\phi'\left(\tau\right)}{a^{2}\left(\tau\right)}
\end{align}
and
\begin{equation}
H=\frac{\dot{a}}{a}=\frac{a'}{a^{2}}=\frac{\mathcal{H}}{a}\Rightarrow \dot{H}=\frac{\mathcal{H}'}{a^{2}}-\frac{\mathcal{H}^{2}}{a^{2}},
\end{equation}
where we have introduced the Hubble parameter in conformal time, $\mathcal{H}= a^\prime/a$. Accordingly, we obtain
\begin{equation}
\dot{H}+2H^{2}=\frac{a''}{a^{3}}.
\end{equation}
In this way, the background field equation, Eq.~\eqref{ddotphi}, becomes
\begin{equation}
\frac{\phi''}{a^{2}}-\mathcal{H}\frac{\phi'}{a^{2}}+3\mathcal{H}\frac{\phi'}{a^{2}}-\frac{\nabla^{2}\phi}{a^{2}}+6\xi\frac{a''}{a^{3}}\phi+V_{,\phi}=0.\label{conformaltimeback}
\end{equation}
that gives
\begin{equation}
\phi''+2\mathcal{H}\phi'-\nabla^{2}\phi+6\xi\frac{a''}{a}\phi=-V_{,\phi}a^{2}. \label{baconf}
\end{equation}
Eq. \eqref{baconf} can be written in the compact form
\begin{equation}
  \frac{1}{\sqrt{-g}}\partial_{\mu}\left(\sqrt{-g}g^{\mu\nu}\partial_{\nu}\phi\right)+V^{\rm eff}_{,\phi}=0,\label{ddotphiN}
\end{equation}
that for the effective potential in Eq. \eqref{effpot} gives
\begin{equation}
\frac{1}{\sqrt{-g}}\partial_{\mu}\left(\sqrt{-g}g^{\mu\nu}\partial_{\nu}\phi\right)+6\xi \frac{a''}{a^{3}}\phi+\chi \phi^{3}-\frac{4\Lambda^{4}}{v^{2}}\phi=0. \label{NewddotphiN}
\end{equation}

The overall variation of Eq.~\eqref{ddotphiN} can be decomposed into four separate components, corresponding to the variations of $\frac{1}{\sqrt{-g}}$, $\sqrt{-g}$, $g^{\mu\nu}$ and $\phi$, respectively. Indeed
\begin{equation}
\begin{split}
    &\delta\left(\frac{1}{\sqrt{-g}}\partial_{\mu}\left(\sqrt{-g}g^{\mu\nu}\partial_{\nu}\phi\right)\right)=\delta\left(\frac{1}{\sqrt{-g}}\right)\sqrt{-g}\partial^{\mu}\partial_{\mu}\phi\\
    &+\delta g^{\mu\nu}\partial_{\mu}\partial_{\nu}\phi
    +\frac{1}{\sqrt{-g}}\left(\delta \sqrt{-g}\right)\partial^{\mu}\partial_{\mu}\phi+\delta \partial_{\mu}\partial^{\mu}\phi,
\end{split}
\end{equation}
and since
\begin{align}
    \delta \left(\frac{1}{\sqrt{-g}}\right)&=-\frac{g_{\mu\nu}\delta g^{\mu\nu}}{2\sqrt{-g}},\\
    \delta \sqrt{-g}&=-g g_{\mu\nu}\frac{\delta g^{\mu\nu}}{2\sqrt{-g}},
\end{align}
we have
\begin{equation}
    \delta\left(\frac{1}{\sqrt{-g}}\partial_{\mu}\left(\sqrt{-g}g^{\mu\nu}\partial_{\nu}\phi\right)\right)=\delta \partial_{\mu}\partial^{\mu}\phi.
\end{equation}
The total variation of Eq.~\eqref{ddotphiN} then gives
\begin{equation}
\begin{split}
    &\delta \partial_{\mu}\partial^{\mu}\phi=\delta \phi''+2\frac{a'}{a}\delta \phi'-\partial_{i}\partial^{i}\delta \phi-2\Psi\phi''-4\frac{a''}{a}\Psi\phi'\\
    &-4\Psi'\phi'=-\xi \delta R \phi a^{2}-\delta \phi V^{\rm eff}_{,\phi\phi}a^{2},
\end{split}
\end{equation}
where the variation of the scalar curvature is
\begin{equation}
\delta R=\frac{1}{a^{2}}\left(2\partial_{i}\partial^{i}\Psi-6\Psi''-24\mathcal{H}\Psi'-12\frac{a''}{a}\Psi\right).
\end{equation}
Using now the field equation for the background field $\phi\equiv \phi_{0}\left(\tau\right)$ in conformal time, i.e.,
\begin{equation}
    \phi''+2\mathcal{H}\phi=-V^{\rm eff}_{,\phi}a^{2},\label{Backfield}
\end{equation}
we write
\begin{equation}
2\Psi \phi''+4\Psi \frac{a'}{a}\phi'=-2\Psi V^{\rm eff}_{,\phi}a^{2},
\end{equation}
finally obtaining
\begin{equation}
\begin{split}
    &\delta \phi''+2\frac{a'}{a}\delta \phi'-\partial_{i}\partial^{i}\delta \phi-4\Psi'\phi'\\
    &=-\xi \delta R \phi a^{2}-\delta\phi V^{\rm eff}_{,\phi\phi}a^{2}-2\Psi V^{\rm eff}_{,\phi}a^{2}.\label{ddotphiNN}
\end{split}
\end{equation}


\section{Quasi-de Sitter inflationary dynamics}\label{AppB}

Inflation is usually modeled assuming vacuum energy domination throughout the slow-roll phase. This translates into a de Sitter scale factor, namely, in cosmic time
\begin{equation}
    a(t)\sim e^{H_It},
\end{equation}
where $H_I$ is the inflationary value of the Hubble constant. Then, in conformal time, we can write
\begin{equation}
  a\left(\tau\right)=-\frac{1}{H_{I}}\frac{1}{\tau}\label{parslow},\ \ \ \ \ \ \tau < 0.
\end{equation}
However, during inflation the Hubble rate is not exactly constant and thus a pure de Sitter evolution would not take into account the slow-roll of the inflaton field and its quantum fluctuations.
More specifically, the inflationary dynamics is better described by a quasi-de Sitter scale factor of the form
\begin{equation}
    a\left(\tau\right)=-\frac{1}{H_{I}}\frac{1}{\tau^{\left(1+m\right)}}\label{parslow},
\end{equation}
with $m\ll 1$, i.e., a small deviation from the pure-de Sitter. Noting that
\begin{equation}
    \epsilon=1-\frac{\mathcal{H}'}{\mathcal{H}^{2}}=1-\frac{1}{1+m}\simeq m,
\end{equation}
we can identify $m$ with the slow-roll parameter $\epsilon$, reobtaining Eq. \eqref{quasiDS}. The condition $\epsilon \ll 1$, which is necessarily verified during slow-roll, ensures that the scale factor remains real and positive throughout the inflationary phase. Moreover, from Eq.~\eqref{parslow}, we get
\begin{align}
    a'&=\frac{\left(1+m\right)}{H_{I}}\tau^{-\left(2+m\right)},\\
    a''&=-\frac{\left(1+m\right)\left(2+m\right)}{H_{I}}\tau^{-(3+m)},
\end{align}
and so
\begin{align}
    \frac{a'}{a}&=-\left(1+m\right)\tau^{-1}\simeq -\left(1+\epsilon\right)\tau^{-1},\label{scalefactor'}\\
    \frac{a''}{a}&=\left(2+3m+m^{2}\right)\tau^{-2}\simeq \left(2+3\epsilon\right)\tau^{-2}.\label{scalefactor''}
\end{align}


\section{Geometric particle production}\label{AppC}

We here discuss  in more detail the geometric mechanism of particle production presented in the main text. Inhomogeneities in the gravitational field are taken into account by introducing the perturbed metric tensor
\be
g_{ab}=g_{ab}^{(0)}+H_{ab}=a^{2}\left(\tau\right)\left(\eta_{ab}+h_{ab}\right),
\ee
where $\eta_{ab}$ is the Minkowski metric and $h_{ab}$ describes perturbations, with $\lvert h_{ab} \rvert \ll 1$. In our framework, metric perturbations are traced back to inflaton fluctuations. Then, during inflation, the interaction Lagrangian is given by Eq. \eqref{intlag}, where $T^{(0)}_{ab}$ is the energy-momentum referred to the fluctuations of the inflaton and it reads explicitly
\begin{equation}
\begin{split}
    T^{(0)}_{ab}&=\partial_{a}\delta\phi\partial_{b}\delta\phi-\frac{1}{2}g_{ab}^{(0)}\left(g^{cd}_{(0)}\partial_{c}\delta\phi\partial_{d}\delta\phi-V\left(\delta \phi\right)\right)\\
    &-\xi\left(\partial_{a}\partial_{b}-g_{ab}^{(0)}\nabla^{c}\nabla_{c}+R^{(0)}_{ab}-\frac{1}{2}R^{(0)}g_{ab}^{(0)}\right)\delta\phi^{2}.\label{EMTPert1}
\end{split}
\end{equation}
The $\hat{S}$-matrix operator relating asymptotic free particle states, `in' ($\tau\rightarrow -\infty$) and  `out'  ($\tau\rightarrow +\infty$), is
\begin{equation}
    \hat{S}=\hat{T}\exp\left[i\int \mathcal{L}_{I}d^{4}x\right],\label{Smatrix0}
\end{equation}
where $\hat{T}$ is the time-ordering operator. Expanding $\hat{S}$ perturbatively in Dyson series up to first order in $h^{ab}$, we get
\begin{equation}
    \hat{S}\simeq 1+i\hat{T}\int \mathcal{L}_{I}d^{4}x\equiv 1+\hat{S}^{(1)}.\label{Smatrix1}
\end{equation}
Then, due to the interaction between metric and the inflaton fluctuations, initial vacuum states evolve into the final state
\begin{equation}
\begin{split}
    &\lim_{\tau\rightarrow +\infty}\left|\Phi\right>=\hat{S}\left|0\right>\\
    &=\left|0\right>+\frac{1}{2}\int d^{3}pd^{3}q {}\braket{p,q|\hat{S}^{(1)}|0}\left|p,q\right>,\label{phiout}
\end{split}
\end{equation}
where
\begin{equation}
    \hat{S}^{(1)}=-\frac{i}{2}\int d^{4}x\sqrt{-g_{(0)}}H^{ab}T^{(0)}_{ab}\label{firstorderSmatrix},
\end{equation}
and
\begin{equation}
    \begin{split}
    &\braket{p,q|\hat{S}^{(1)}|0}=\braket{p,q|i\hat{T}\int d^{4}x \mathcal{L}_{I}|0}, \label{firstorderSelement}
    \end{split}
\end{equation}
are the first order $\hat{S}$-matrix and the $\hat{S}$-matrix element, respectively.

However, in Eq. \eqref{phiout} we have not taken into account that in curved spacetime 'in' and 'out' states are generally different, due to the universe evolution. This implies that an initial vacuum state is no longer seen as a vacuum in the 'out' region. Accordingly, creation and annihilation operators are not the same in the two regions: introducing the ladder operators $b_p$ and $b_p^\dagger$ in the 'out' region, we can write $a_{p}\left|0\right>_{in}=b_{p}\left|0\right>_{out}=0$.

Ladder operators in the two regions are connected by the Bogoliubov transformation
\begin{equation}
b_{p}=\alpha_{p}a_{p}+\beta^{*}_{p}a^{\dag}_{-p},\label{Bogol}
\end{equation}
where $\alpha_p$ and $\beta_p$ are known as Bogoliubov coefficients and satisfy the normalization condition
\begin{equation}
    \left|\alpha_{p}\right|^{2}-\left|\beta_{p}\right|^{2}=1.\label{normaliz}
\end{equation}
Including Bogoliubov transformations in our particle production estimate implies that we have to compute the expectation value of the number operator $N=\frac{1}{\left(2\pi a\right)^{3}}\int d^{3}k b^{\dag}_{k}b_{k}$ in the final state, i.e.,
\begin{equation}
    \braket{\Phi|N|\Phi}=N^{(0)}+N^{(1)}+N^{(2)}\label{sumthreeterms}.
\end{equation}
The first term $N^{(0)}$ denotes the creation rate due to the background only. Indeed, the homogeneous expansion combines modes of positive and negative frequency, so there exists some values of $p$ for which $\beta_{p}\neq0$ in Eq.~\eqref{Bogol}, leading to the creation of particles. This is usually known as \emph{gravitational particle production}. The average number density of created particles at zero order, with proper normalization, is
\begin{equation}
    N^{(0)}=\frac{1}{\left(2\pi a\right)^{3}}\int d^{3}p \braket{0|b^{\dag}_{p}b_{p}|0}=\frac{1}{\left(2\pi a\right)^{3}}\int d^{3}k \left|\beta_{p}\right|^{2}.\label{N0}
\end{equation}
The second term $N^{(1)}$ is the result of the combined effects due to interaction and background, i.e., it arises from the interference between 0-particle and 2-particle states and it is given by
\begin{equation}
\begin{split}
    &N^{(1)}=\frac{1}{\left(2\pi a\right)^{3}}\int d^{3}p d^{3}q\delta^{3}\left({\bf p}+{\bf q}\right)\\
    &\times{\rm Re}\left[\braket{p,q|\hat{S}^{(1)}|0}\left(\alpha_{p}\beta_{p}+\alpha_{q}\beta_{q}\right)\right].\label{N1}
\end{split}
\end{equation}

Finally, the last term $N^{(2)}$ arises from 2-particle states only and reads
\begin{equation}
    N^{(2)}=\frac{1}{\left(2\pi a\right)^{3}}\int d^{3}p d^{3}q \left|\braket{0|\hat{S}|p,q}\right|^{2}\left(1+\left|\beta_{p}\right|^{2}+\left|\beta_{q}\right|^{2}\right).\label{N2}
\end{equation}

We notice that, starting from second order in the inhomogeneities, pair production is no longer restricted to particle-antiparticle pairs, which in principle may annihilate.

This is related to the presence of inhomogeneities, which break space translation symmetry so that momentum conservation does not apply at this stage. In this work, we focused on the the computation of Eq. \eqref{N2}, setting to zero the Bogoliubov coefficients as a first estimate.

This choice gives
\begin{equation} \label{N2geom}
    N^{(2)}=\frac{1}{\left(2\pi a\right)^{3}}\int d^{3}p d^{3}q \left|\braket{0|\hat{S}|p,q}\right|^{2},
\end{equation}
where the `in' and `out' vacua are the same, since the Bogoliubov coefficients $\beta_{p,q}$ are set to zero.

However, we see that the presence of non-zero Bogoliubov coefficients is not only responsible for gravitational particle production at zero and first order, but also enhances geometric production at second perturbative order. For this reason, a more refined estimate of the total number of produced particles will require the inclusion of such coefficients in our calculations, which will be object of future efforts.


\section{Model dependence on the cut-off time value}\label{AppD}

As discussed in Sec. \ref{sezione3}, we focus on particle production for super-Hubble modes $k < a H_I$. At the same time, in order to allow a causal generation mechanism for fluctuations, we also assume that all the modes of interest are sub-Hubble at the beginning of inflation. For this reason, we need to introduce a cut-off time $\tau^\prime$, so to evaluate particle production within the time interval $\tau \in \left[ \tau^\prime, \tau_f \right]$, where the modes $a(\tau_i) H_I < k < a(\tau^\prime) H_I$ can be expressed in super-Hubble form.  Thus, passing to spherical coordinates, the number density of Eq.~\eqref{N2geom} at the end of inflation reads
\begin{equation}
\begin{split}
&N^{\left(2\right)}\left(\tau_{f}\right)=\frac{a^{-3}\left(\tau_{f}\right)}{\left(2\pi\right)^{3}}\int_{a\left(\tau_{i}\right)H_{I}}^{a\left(\tau'\right)H_{I}}\int _{0}^{2\pi} \int _{0}^{2\pi} \int _{0}^{\pi} \int _{0}^{\pi}\\
&dqdpd\theta d\sigma d\varphi d\rho q^{2}p^{2}\sin\theta\sin\sigma \left|\braket{0|\hat{S}|p,q}\right|^{2},\label{NP1}
\end{split}
\end{equation}
where the Bogoliubov coefficients are set to zero, i.e., particles produced due to the sole expansion of the universe are not analyzed here. In Eq.~\eqref{NP1}, $\theta,\sigma \in \left[0,\pi\right]$ and $\varphi,\rho \in \left[0,2\pi\right]$, with $q, p$ defined as $q=\sqrt{q_{x}^{2}+q_{y}^{2}+q_{z}^{2}}$
and $p=\sqrt{p_{x}^{2}+p_{y}^{2}+p_{z}^{2}}$. At this point, it is evident from Eq.~\eqref{NP1} that the cut-off time $\tau'$ can play a significant role in the quantity of particles produced. In general, any choice $\tau^\prime > \tau_i$, where $\tau_i$ denotes the initial time for inflation, inevitably leads to underestimate the total number of geometric particles produced. However, as noted elsewhere \cite{CCPBelfiglio0}, the contribution of geometric particle production is typically enhanced at small momenta for bosonic fields, in analogy to nonperturbative gravitational particle production. This implies that the contribution of modes $k \gg a(\tau_i)H_I$ to the number density is typically negligible\footnote{The same conclusion is not necessarily true in case of fermionic fields, since the mode dependence of the number density for produced particles can be significantly influenced by the statistics of the field involved.}. However, a more refined estimate of geometric particle production would necessarily include the contribution of all modes that cross the horizon after $\tau^\prime$. For the moment, we show below the number density obtained for different values of $\tau'$, in both small and large field domains.

\subsection{Small-field domain}\label{AppD1}
\begin{figure}[H]
\centering
        \includegraphics[width=8cm]{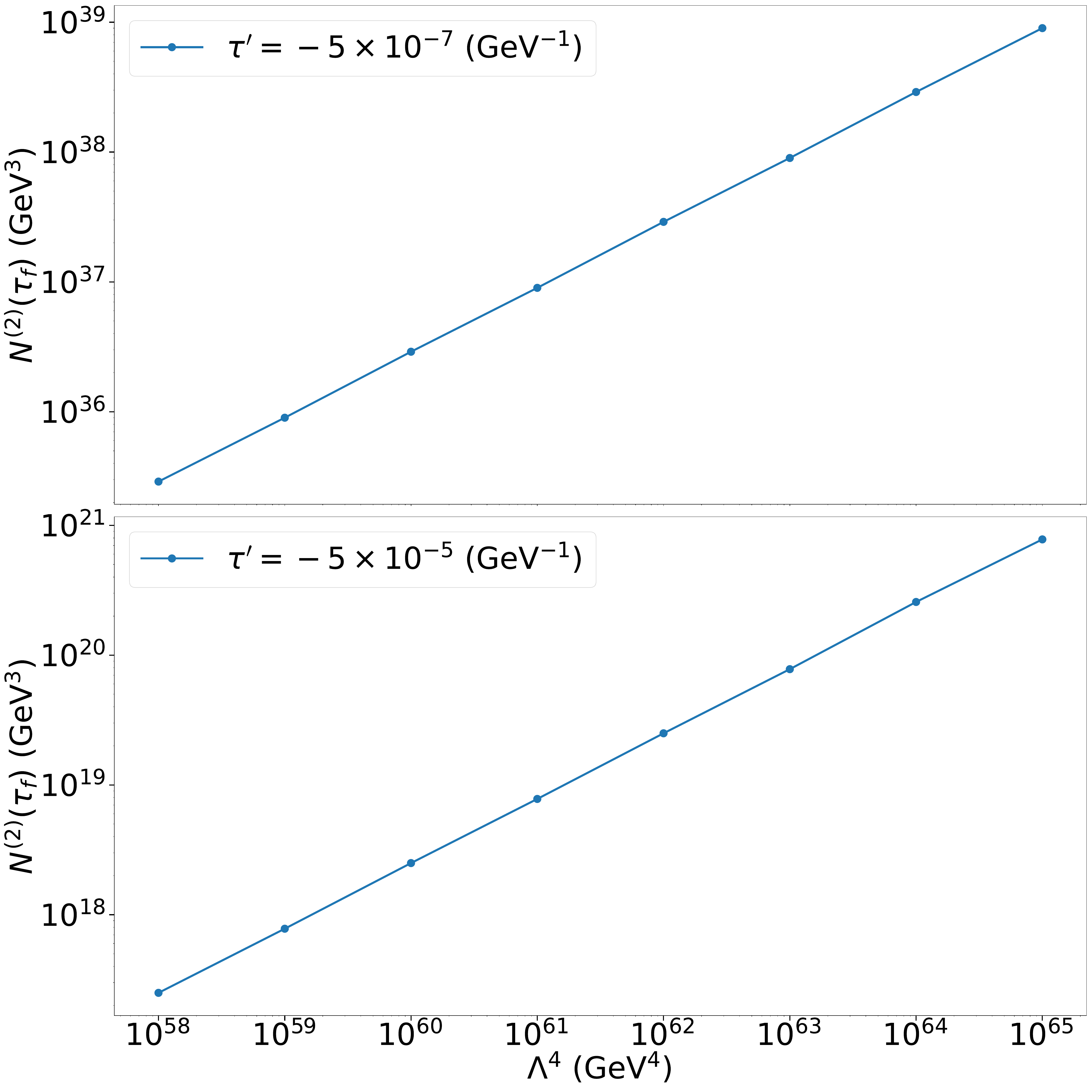}
    \caption{Number density $N^{(2)}$ in $\unit{\giga\electronvolt}^{3}$ as function of vacuum energy $\Lambda^{4}$ in loglog scale. The number density is computed assuming $N=60$ e-foldings, $\tau_{i}=-10^{3}$ \unit{\giga\electronvolt}$^{-1}$, $\epsilon=10^{-3}$ and $\xi=10^{-5}$. The cut-off time is chosen as $\tau'=-5\times 10^{-7}$ \unit{\giga\electronvolt}$^{-1}$ and $\tau'=-5\times 10^{-5}$ \unit{\giga\electronvolt}$^{-1}$, respectively.}
    \label{VarLsmalldifftau}
\end{figure}

\begin{table}[H]
\begin{center}
\begin{tabular}{||c | c | c | c||}
\hline
$\tau^\prime$ (\unit{\giga\electronvolt})$^{-1}$ & $\Lambda^{4}$ (\unit{\giga\electronvolt})$^{4}$  & $\chi$ & $N^{(2)}\left(\tau_{f}\right)$
 (\unit{\giga\electronvolt})$^{3}$\\ [0.5ex]
 \hline\hline
& $\num{1.0}\times 10^{58}$ & $1.0\times 10^{-20}$ & $\num{2.9}\times 10^{35}$  \\
& $\num{1.0}\times 10^{59}$& $1.0\times 10^{-19}$ &  $\num{9.0}\times 10^{35}$ \\
& $\num{1.0}\times 10^{60}$ & $1.0\times 10^{-18}$ & $\num{2.9}\times 10^{36}$ \\
$-5\times 10^{-7}$ & $\num{1.0}\times 10^{61}$ & $1.0\times 10^{-17}$ & $\num{9.0}\times 10^{36}$ \\
& $\num{1.0}\times 10^{62}$ & $1.0\times 10^{-16}$ & $\num{2.9}\times 10^{37}$ \\
& $\num{1.0}\times 10^{63}$ & $1.0\times 10^{-15}$ & $\num{9.0}\times 10^{37}$ \\
& $\num{1.0}\times 10^{64}$   & $1.0\times 10^{-14}$ & $\num{2.9}\times 10^{38}$ \\
& $\num{1.0}\times 10^{65}$ & $1.0\times 10^{-13}$ & $\num{9.0}\times 10^{38}$ \\
\hline
& $\num{1.0}\times 10^{58}$ & $1.0\times 10^{-20}$ & $\num{2.5}\times 10^{17}$  \\
& $\num{1.0}\times 10^{59}$& $1.0\times 10^{-19}$ &  $\num{7.8}\times 10^{17}$ \\
& $\num{1.0}\times 10^{60}$ & $1.0\times 10^{-18}$ & $\num{2.5}\times 10^{18}$ \\
$-5\times 10^{-5}$ & $\num{1.0}\times 10^{61}$ & $1.0\times 10^{-17}$ & $\num{7.8}\times 10^{18}$ \\
& $\num{1.0}\times 10^{62}$ & $1.0\times 10^{-16}$ & $\num{2.5}\times 10^{19}$ \\
& $\num{1.0}\times 10^{63}$ & $1.0\times 10^{-15}$ & $\num{7.8}\times 10^{19}$ \\
& $\num{1.0}\times 10^{64}$   & $1.0\times 10^{-14}$ & $\num{2.5}\times 10^{20}$ \\
& $\num{1.0}\times 10^{65}$ & $1.0\times 10^{-13}$ & $\num{7.8}\times 10^{20}$ \\
\hline
\end{tabular}
\end{center}
\caption{Number density of geometric particles produced in the small-field scenario, for different values of vacuum energy and corresponding self-coupling constant.}
\label{Tab1app}
\end{table}

\subsection{Large-field domain}\label{AppD2}

\begin{figure}[H]
        \includegraphics[width=8cm]{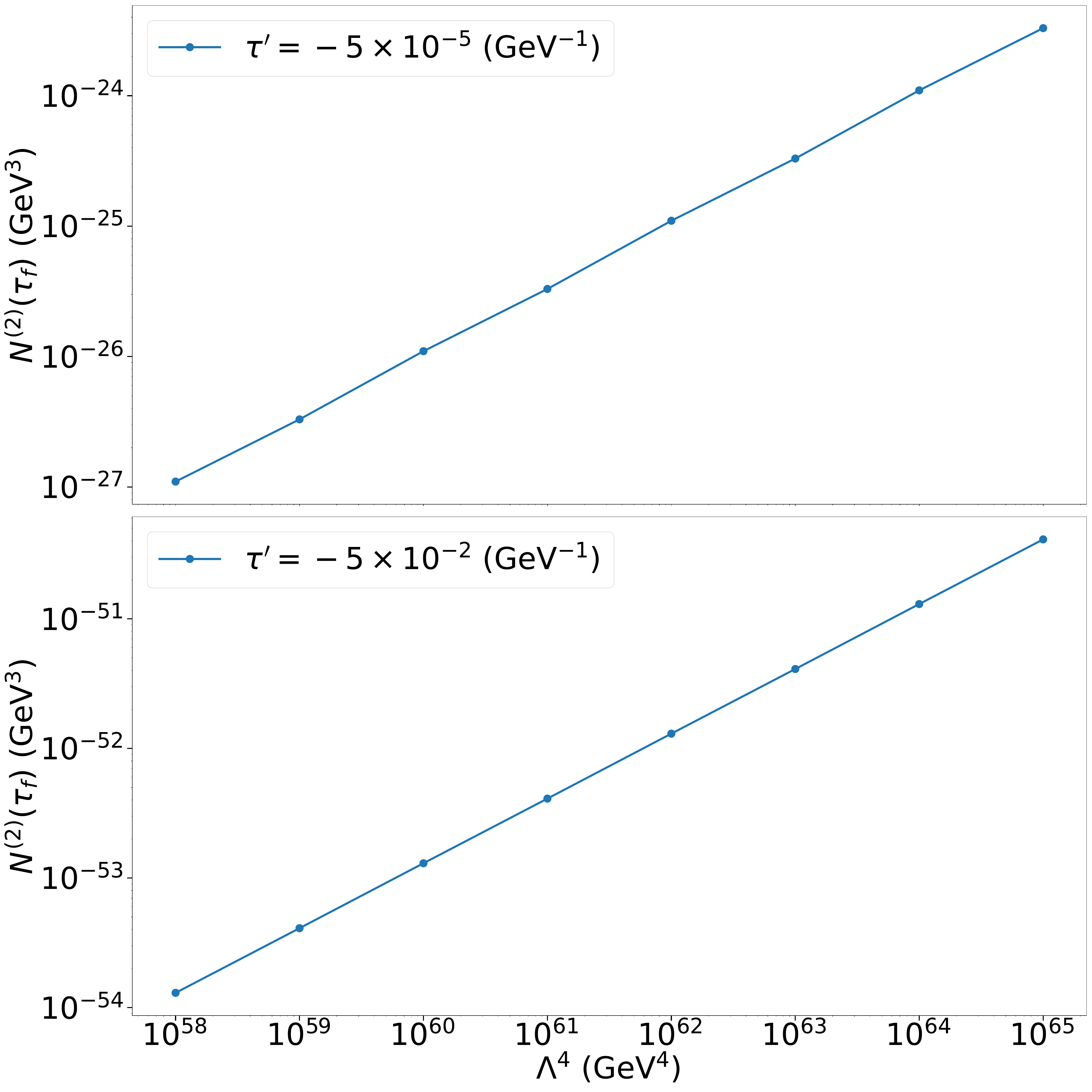}
    \caption{Number density $N^{(2)}$ in $\unit{\giga\electronvolt}^{3}$ as function of vacuum energy $\Lambda^{4}$ in loglog scale. The number density is computed assuming $N=60$ e-foldings, $\tau_{i}=-10^{3}$ \unit{\giga\electronvolt}$^{-1}$, $\epsilon=10^{-3}$ and $\xi=10^{-3}$. The cut-off time is chosen as $\tau'=-5\times 10^{-5}$ \unit{\giga\electronvolt}$^{-1}$ and $\tau'=-5\times 10^{-2}$ \unit{\giga\electronvolt}$^{-1}$, respectively.}
    \label{VarLlarge2}
\end{figure}

\begin{table}[H]
\begin{center}
\begin{tabular}{||c | c | c | c||}
\hline
 $\tau^\prime$ (\unit{\giga\electronvolt})$^{-1}$ &$\Lambda^{4}$ (\unit{\giga\electronvolt})$^{4}$ & $\chi$ & $N^{(2)}\left(\tau_{f}\right)$
 (\unit{\giga\electronvolt})$^{3}$\\ [0.5ex]
 \hline\hline
 & $\num{1.0}\times 10^{58}$ & $\num{3.2}\times 10^{-21}$  & $\num{1.1}\times 10^{-27}$  \\
& $\num{1.0}\times 10^{59}$&  $\num{3.2}\times 10^{-20}$ & $\num{3.3}\times 10^{-27}$ \\
&$\num{1.0}\times 10^{60}$  &  $\num{3.2}\times 10^{-19}$  & $\num{1.1}\times 10^{-26}$\\
$-5\times 10^{-5}$ &$\num{1.0}\times 10^{61}$ & $\num{3.2}\times 10^{-18}$  & $\num{3.3}\times 10^{-26}$\\
& $\num{1.0}\times 10^{62}$ & $\num{3.2}\times 10^{-17}$  & $\num{1.1}\times 10^{-25}$ \\
& $\num{1.0}\times 10^{63}$ & $\num{3.2}\times 10^{-16}$  & $\num{3.3}\times 10^{-25}$ \\
& $\num{1.0}\times 10^{64}$   & $\num{3.2}\times 10^{-15}$  & $\num{1.1}\times 10^{-24}$ \\
& $\num{1.0}\times 10^{65}$ & $\num{3.2}\times 10^{-14}$ & $\num{3.3}\times 10^{-24}$ \\
\hline
& $\num{1.0}\times 10^{58}$ & $\num{3.2}\times 10^{-21}$  & $\num{1.3}\times 10^{-54}$  \\
& $\num{1.0}\times 10^{59}$&  $\num{3.2}\times 10^{-20}$ & $\num{4.1}\times 10^{-54}$ \\
&$\num{1.0}\times 10^{60}$  &  $\num{3.2}\times 10^{-19}$  & $\num{1.3}\times 10^{-53}$\\
$-5\times 10^{-2}$ &$\num{1.0}\times 10^{61}$ & $\num{3.2}\times 10^{-18}$  & $\num{4.1}\times 10^{-53}$\\
& $\num{1.0}\times 10^{62}$ & $\num{3.2}\times 10^{-17}$  & $\num{1.3}\times 10^{-52}$ \\
& $\num{1.0}\times 10^{63}$ & $\num{3.2}\times 10^{-16}$  & $\num{4.1}\times 10^{-52}$ \\
& $\num{1.0}\times 10^{64}$   & $\num{3.2}\times 10^{-15}$  & $\num{1.3}\times 10^{-51}$ \\
& $\num{1.0}\times 10^{65}$ & $\num{3.2}\times 10^{-14}$ & $\num{4.1}\times 10^{-51}$ \\
\hline
\end{tabular}
\end{center}

\caption{Number density of particles produced at different cut-off times, for given values of vacuum energy and corresponding self-coupling constant.}
\label{Tab2'}
\end{table}

\end{document}